\documentclass[]{emulateapj}
\usepackage{natbib,amssymb}
\newcommand{\mic}{ $\mu$m }

\newcommand{\degree}{\ensuremath{^\circ} }
\newcommand{\ione}{[3.6$\mu$]}
\newcommand{\itwo}{[4.5$\mu$]}
\newcommand{\ithree}{[5.8$\mu$]}
\newcommand{\ifour}{[8$\mu$]}
\bibpunct{(}{)}{;}{a}{}{,}
\usepackage{graphics,graphicx,subfigure,multirow}
\begin{document}

\title{Lifting the Dusty Veil II: \\ A Large-Scale Study of the Galactic Infrared Extinction Law}
\shorttitle{Galactic Infrared Extinction Law}
\shortauthors{Zasowski et al.}
\journalinfo{ApJ}
\submitted{Accepted to ApJ}

\author{G.~Zasowski\altaffilmark{1}, S.~R.~Majewski\altaffilmark{1}, R.~Indebetouw\altaffilmark{1}, M.~R.~Meade\altaffilmark{2}, D.~L.~Nidever\altaffilmark{1},  \\
R.~J.~Patterson\altaffilmark{1}, B.~Babler\altaffilmark{2}, M.~F.~Skrutskie\altaffilmark{1}, C.~Watson\altaffilmark{3}, 
B.~A.~Whitney\altaffilmark{2,4}, E.~Churchwell\altaffilmark{2}} 

\altaffiltext{1}{Department of Astronomy, University of Virginia, PO Box 400325, Charlottesville, VA 22904-4325: gailis@virginia.edu, srm4n@virginia.edu, remy@virginia.edu, dln5q@virginia.edu, rjp0i@virginia.edu, mfs4n@virginia.edu}
\altaffiltext{2}{Department of Astronomy, University of Wisconsin at Madison, 475 North Charter St., Madison, WI 53706: meade@astro.wisc.edu, brian@sal.wisc.edu, bwhitney@spacescience.org, ebc@astro.wisc.edu}
\altaffiltext{3}{Manchester College, North Manchester, IN 46962: cwatson@manchester.edu}
\altaffiltext{4}{Space Science Institute, 4750 Walnut St., Boulder, CO 80301}

\begin{abstract}
We combine near-infrared (2MASS) and mid-infrared ({\it Spitzer}-IRAC) photometry to characterize the IR extinction law (1.2-8 $\mu$m) 
over nearly 150$^\circ$ of contiguous Milky Way midplane longitude.
The relative extinctions in 5 passbands across these wavelength and longitude ranges are derived 
by calculating color excess ratios for G and K giant red clump stars in contiguous midplane regions and deriving the wavelength dependence of extinction in each one.
Strong, monotonic variations in the extinction law shape are found as a function of angle from the Galactic center, symmetric on either side of it.
These longitudinal variations persist even when dense interstellar regions, known a priori to have a shallower extinction curve, are removed.  
The increasingly steep extinction curves towards the outer Galaxy indicate a steady decrease in the 
absolute-to-selective extinction ratio ($R_V$) and in the mean dust grain size at greater Galactocentric angles.
We note an increasing strength of the 8\mic extinction inflection at high Galactocentric angles and,
using theoretical dust models, show that this behavior is consistent with the trend in $R_V$.
Along several lines of sight where the solution is most feasible,
$A_\lambda$/$A_{Ks}$ as a function of Galactic radius ($R_{GC}$) is estimated and shown to have a Galactic radial dependence.
Our analyses suggest that the observed relationship between extinction curve shape and 
Galactic longitude is due to an intrinsic dependence of the extinction law on Galactocentric radius.
\end{abstract}

\keywords{dust, extinction --- Galaxy: disk --- infrared: ISM}


\section{INTRODUCTION}

Interstellar extinction presents one of the longest-standing challenges to observational astronomy.
Extensive work has been done to understand the complex
and highly-variable ultraviolet and optical portions of the extinction curve (see, e.g., reviews by \citealt{Mathis_90_dustreview} and \citealt{Whittet_03_dust}); 
only recently, however, has comprehensive characterization of the infrared (IR) extinction law become possible, through the influx of near- and mid-infrared (NIR, MIR)
data provided by large-area surveys like 2MASS, UKIDSS, and GLIMPSE I/II/3D.
Not only does this characterization enable more accurate study of sources extinguished by dust,
but reliable derivation of the wavelength-dependent IR extinction behavior also facilitates identification and modeling of the physical properties
(e.g., chemical composition, grain size, crystallization fraction) of the interstellar dust itself.

Many dust absorption studies to date have converged to a treatment of the infrared extinction law in the diffuse interstellar medium (ISM) as nearly constant and universal. 
The NIR extinction curve ($\lambda$ $\lesssim$ 1--4 $\mu$m) 
is often modeled as a power law ($A_\lambda\propto\lambda^{-\beta}$), with $\beta$ ranging
between 1.6 and 1.8 \citep[e.g.,][and references therein]{RiekeLeb_85_extlaw,Draine_03_dust}, but values up to 2 have been reported \citep{Nishi_06_extlaw,Nishi_09_extlaw}.
At longer wavelengths ($\lambda$ $>$ 4 $\mu$m), however, $A_\lambda$ deviates from this
relationship and appears to ``flatten out'' (become grayer) before increasing towards the peak of the 9.7\mic silicate feature.  
The absolute extinction ratios $A_{\lambda_0}$/$A_{Ks}$ used to normalize the derived extinction curves are generally 
determined for the diffuse ISM using reddened background standard
candles for which intrinsic colors can be assumed---e.g., red clump stars \citep{Indeb_05_extlaw,Nishi_09_extlaw} or RGB-tip or low-mass-loss AGB stars
\citep{Jiang_03_extlaw,Jiang_06_7-15ext}.
 
More recent work has demonstrated that in regions of dense ISM, such as dark cloud cores or star-forming regions, 
the MIR $A_\lambda$ curve becomes even shallower than in diffuse material, 
likely due to dust grain growth through grain coagulation or ice mantle coating 
\citep[e.g.,][]{WeinDraine_01_dustsize,Moore_05_extlaw,RZ_07_extlaw,Flaherty_07_extlaw,McClure_08_extlaw,Chapman_08_extlaw}.  
The absolute ratios $A_{\lambda_0}$/$A_{Ks}$ have been determined in these regions using either stars assumed to be in the background \citep{Flaherty_07_extlaw} 
or associated gas emission---e.g., hydrogen recombination spectra \citep{Lutz_96_galcenter,Moore_05_extlaw} or 
H$_2$ rotational and rovibrational lines \citep{Rosenthal_00_H2extlaw}.  Infrared extinction behavior as a function
of density, beyond a simple ``dense'' vs. ``diffuse'' dichotomy, has only recently begun to be studied in 
specific cloud cores \citep{Chapman_08_extlaw,McClure_08_extlaw}. 
We address the effects of ISM density on the extinction law here as well, but the present work is also the first to 
examine extinction behavior with respect to overall Galactic ISM density or grain-size gradients.

Given the differences in the extinction law behavior in dense and diffuse interstellar environments, 
one may expect the Galactic extinction law to vary substantially through the widely changing environments across the disk.
Yet the vast majority of infrared extinction law studies, including those listed above, use no more than a few specific regions on the sky ($\lesssim$10 deg$^2$) 
to define the extinction law and/or are limited to measuring relative $A_\lambda$ variations in environmental extremes, such as dark clouds.
Moreover, most efforts are concentrated in the inner Galaxy, where both starcounts and the overall extinction are higher, but where the potential 
effects of Galactic-scale gradients (such as ISM metallicity or density) cannot be easily identified.
Thus, there is little reason for the ``universal'' extinction law derived from these observations to be in fact applicable to the Galaxy as a whole.

In contrast to previous studies, this work tests both the very assumption of a universal diffuse extinction law 
and the pertinence of a simple bimodal {\it dense} versus {\it diffuse} description of the ISM with regard to extinction.  
Our collected databases allow us to determine coherent patterns of extinction behavior by measuring the IR extinction law in a 
consistent way along many different sightlines spread over $\sim$150$^{\circ}$ of the Galactic disk.  
\S\ref{sec:data} of this paper contains a description of the data we use in this new, comprehensive exploration of extinction behavior.  
\S\ref{sec:measure_ext} explains our methodology for determining $A_\lambda$/$A_{Ks}$, 
and \S\ref{sec:density} examines the question of density as the sole driver of extinction variations.
In \S\ref{sec:discuss}, we compare our results to previous observational and theoretical studies and
discuss their similarities and differences with our findings.
We also address the implications of the observed variations in the 8\mic extinction upturn and the feasibility of measuring extinction behavior as a function of
Galactocentric radius.  Finally, \S\ref{sec:conc} summarizes our findings and conclusions.

\section{DESCRIPTION OF DATA} \label{sec:data}

To probe the extinction law from 1.2--8\mic throughout a large portion of the Galactic disk, we combine photometric data from 
several mid-infrared {\it Spitzer}-IRAC surveys with the near-infrared 2MASS Point Source Catalog \citep[PSC;][]{Skrutskie_06_2mass}.
The PSC contains photometric measurements at {\it J} (1.24 $\mu$m), {\it H} (1.66 $\mu$m), and {\it K$_s$} (2.16 $\mu$m), and is considered complete
to a magnitude limit of 16.1, 15.5, and 15.1 mag in
$J$, $H$, and $K_s$, respectively.\footnote{http://www.ipac.caltech.edu/2mass/releases/allsky/doc/sec6\_5a1.html}

At MIR wavelengths, we use data taken with {\it Spitzer}'s IRAC instrument, with imaging channels roughly centered at 3.6, 4.5, 5.8, and 8 $\mu$m.  
From the {\it Spitzer} Legacy archive, we have used photometry from the GLIMPSE-I catalog \citep{Benjamin_03_glimpse}, which consists of 2-sec IRAC 
observations of the Galactic plane, contiguously mapped for $|b|$ $\leq$ 1$^\circ$ across 10\degree $\leq$ $l$ $\leq$ 65\degree and 295\degree $\leq$ $l$ $\leq$ 350\degree.

We supplement the GLIMPSE coverage with our own Galactic disk observations.  
Extending the reach of GLIMPSE is a similar survey of the Galactic midplane covering the Vela molecular ridge and the Vela-Carina star-forming
region (PID 40791).  The survey region stretches between 255\degree $\leq$ $l$ $\leq$ 295\degree and is 2\degree tall in latitude, 
either centered at the midplane or offset $\pm$0.5$^\circ$ to include regions of interest or the expected disk warp at these Galactic longitudes;
the depth of the photometry is very similar to that of GLIMPSE (e.g., $m_{3.6\mu m}$ $\lesssim$ 15.5). 
Additional IRAC observations of the disk include a number of 1$^{\circ}$$\times$1$^{\circ}$ fields scattered unevenly between 
90\degree $\leq$ $l$ $\leq$ 300\degree and 0\degree $\leq$ $|b|$ $\leq$ 12\degree (the ``Argo'' survey, PID 20499).

\begin{figure*}[htb!]
 \begin{center}
 \includegraphics[angle=90,width=0.8\textwidth]{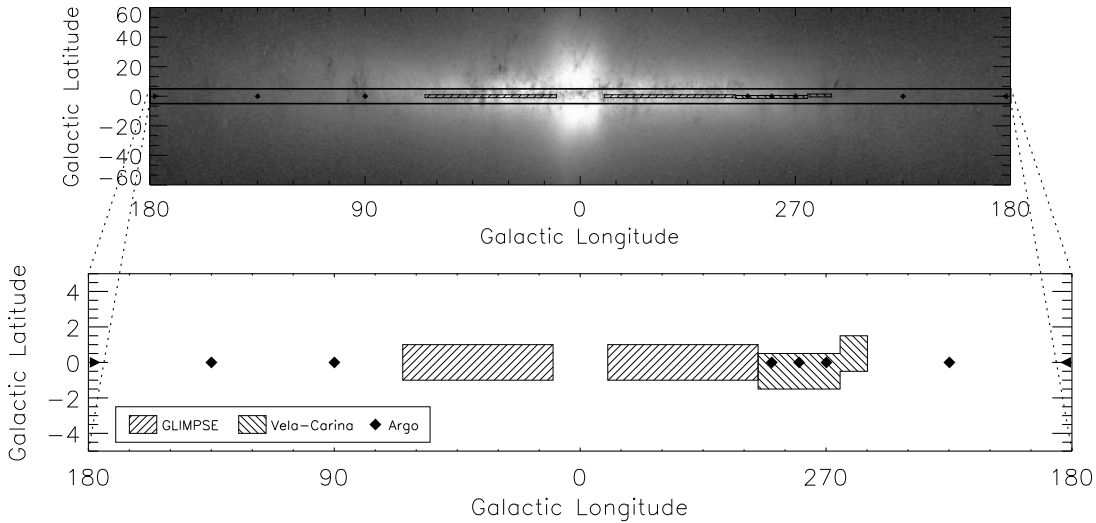}
 \end{center}
 \caption{{\it Top:} 2MASS all-sky image; horizontal lines indicate $\pm$ 5$^\circ$ latitude. {\it Bottom:} Close-up of Galactic midplane. 
The overlaid surveys on both plots are GLIMPSE (forward-slash shading), Vela-Carina (backward-slash shading), and Argo (filled diamonds), 
as detailed in Table~\ref{tab:data}. 
For the Argo data, we have only shown the midplane fields used in this study (i.e., with $|b|$ $<$ 1$^\circ$), 
and we have slightly elongated the points in longitude for visual clarity (the fields are actually 1$^\circ$x1$^\circ$ each).}
 \label{fig:datamap}
\end{figure*}

All of these {\it Spitzer}-IRAC data have been reduced and band-merged with the 2MASS PSC
using the same pipeline\footnote{http://irsa.ipac.caltech.edu/data/SPITZER/GLIMPSE/doc/glimpse1\_dataprod\_v2.0.pdf}, and
we begin with the more complete Archive data for all of the IRAC surveys.
In total, we have data in 7 photometric bands (2MASS {\it JHK$_s$} and IRAC \ione, \itwo, \ithree, \ifour) 
spanning $\sim$150$^{\circ}$ of disk longitude ($\sim$290 deg$^2$ in total area; see Figure~\ref{fig:datamap}), observed and reduced in a consistent manner.  
Table~\ref{tab:data} contains the sky coverage and references for the IRAC observations.  

\begin{deluxetable*}{c c c c}[htpb] 
\tablewidth{0pt}
\tablecaption{Span of IRAC observations used in this analysis.} 
\tablehead{\colhead{Survey} & \colhead{Galactic Longitude, $l$} & \colhead{Galactic Latitude, $b$} & \colhead{Reference}}
\startdata
\multirow{2}{*} {GLIMPSE} & 10 - 65 & \multirow{2}{*} {(-1) - +1} & \multirow{2}{*} {\citet{Benjamin_03_glimpse}} \\
& 295 - 350 & & \\
Vela-Carina & 255 - 295 & (-1.5) - +1.5\tablenotemark{a} & {\it Spitzer} PID 40791 \\
\multirow{2}{*} {Argo} & 90, 135, 180, & \multirow{2}{*} {(-0.5) - (0.5)} & \multirow{2}{*} {{\it Spitzer} PID 20499} \\
& 225, 270, 280, 290 & & \\
\enddata
\tablenotetext{a}{Survey spans 2\degree at any given location; center ranges from -0.5 $\leq$ $b$ $\leq$ 0.5.}
\label{tab:data}
\end{deluxetable*}

\section{MEASUREMENT OF EXTINCTION} \label{sec:measure_ext}

We use the method of color excess ratios to measure extinction behavior as a function of wavelength.  This procedure consists of making linear fits to pairs
of observed colors for numerous stars with varying levels of extinction, as described in detail in Section~\ref{sec:CERfit}.  
To minimize intrinsic scatter in the stellar colors, which would make the fits less 
certain and less physically meaningful, we desire a sample of stars with homogeneous stellar atmospheric properties; 
we also require the adopted stellar population be common enough in the disk to comprise a significant statistical sample.  
Section~\ref{sec:sample} contains the details of our sample selection criteria.

\subsection{Sample Selection} \label{sec:sample}
To ensure the most homogeneous sample of stars for study, we choose to use G and K type red clump (RC) stars, a population that is relatively
luminous \citep[$M_{Ks}$ $\sim$ $-$1.54;][]{Groenewegen_08_RCmag}, homogeneous in color and absolute magnitude 
\citep[($J-K_s$) $\sim$ 0.65 $\pm$ 0.10, $\sigma_{M_{Ks}}$ $\sim$ 0.04;][]{Girardi_02_isochrones,Groenewegen_08_RCmag},
and common in disk populations across a wide range of metallicities and ages.  In the color-magnitude diagram (CMD) of an unreddened region of sky, the RC
appears as a prominent vertical stripe, with the counts per magnitude tracing the density of the stellar population along the line of sight.  
When dust is present along the line of sight, 
its extinction acts to both dim the RC stars and redden them, resulting in a diagonal shift of their CMD to the right at fainter magnitudes (Figure~\ref{fig:cmd}a).  
However, even in highly-reddened disk midplane fields, the RC
forms a distinctive locus, which we utilize to identify the stars for study.  

\begin{figure}[htpb]
 \begin{center}
 \includegraphics[angle=90,width=0.5\textwidth]{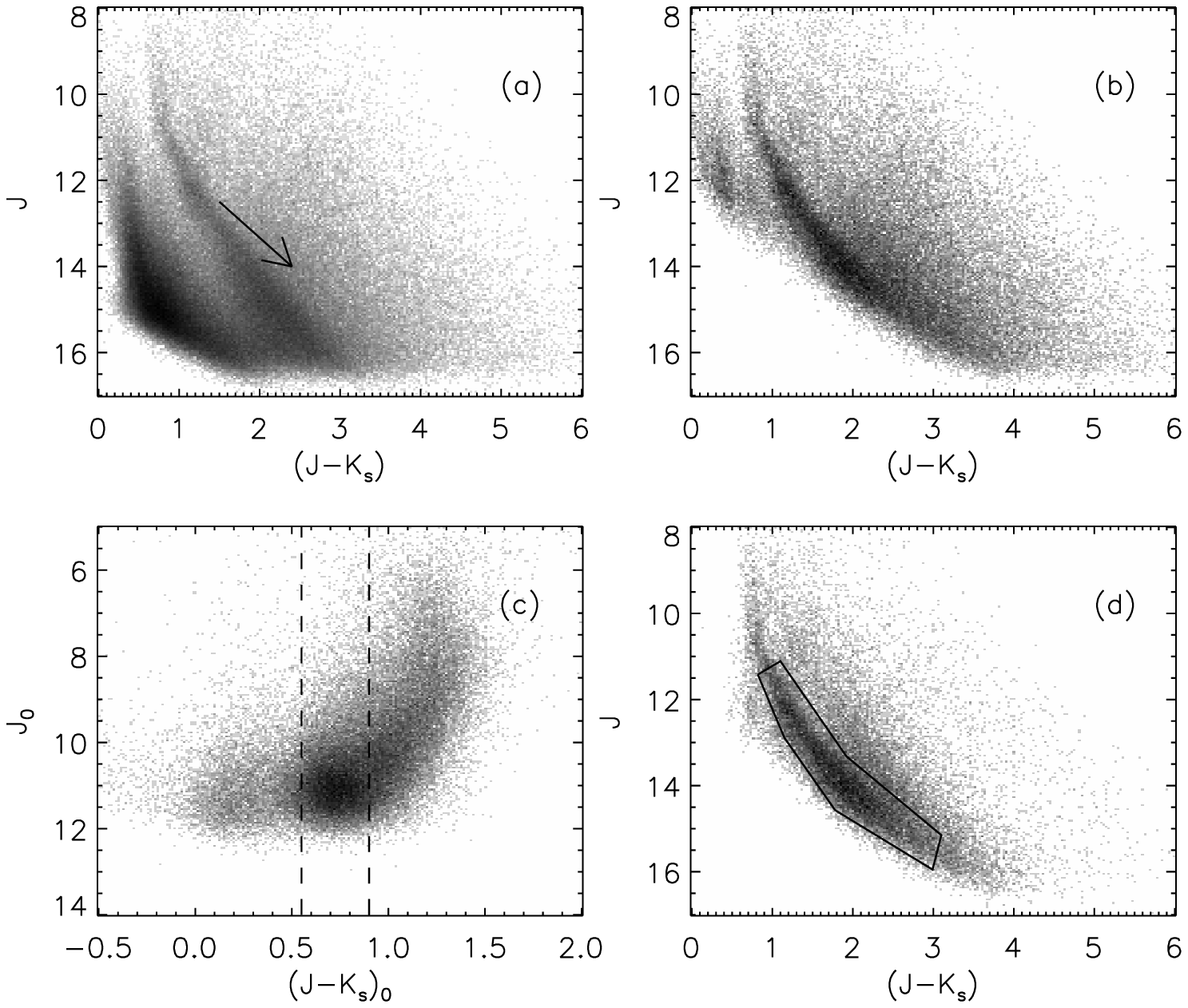}
 \end{center}
 \caption{Color-magnitude diagrams (CMDs) demonstrating red clump (RC) sample selection for a typical field, centered at ($l$=310$^\circ$, $b$=0$^\circ$).  
{\it Panel~(a)}:~the near-infrared CMD with all stars shown; the RC locus is clearly visible next to the arrow indicating a typical reddening vector \citep[$A_{Ks}$ = 1.5 mag;][]{Indeb_05_extlaw}.  
{\it Panel~(b)}:~CMD of the stars remaining after photometric uncertainty and color limits are imposed.  
{\it Panel~(c)}:~dereddened CMD, with main-sequence, RC, and red giant branch stars visible; the dotted lines indicate the color range used to remove the dwarfs and red giant stars.  Notice the magnified $(J-K_s)_0$ scale in this panel.
{\it Panel~(d)}:~CMD of stars remaining after the dereddened-color cut; the boxed region indicates the final RC selection.}
 \label{fig:cmd}
\end{figure}

For our analysis, we selected only those stars detected in all three of the 2MASS {\it JHK$_s$} bands with $\sigma$ $\leq$ 0.4 mag; 
this requirement removes many intrinsically red sources (i.e., those not detected in the bluer 2MASS bands) 
with low line-of-sight reddening from potentially contaminating the highly-reddened part of the RC sample.
For the four {\it Spitzer}-IRAC bands (\ione, \itwo, \ithree, \ifour), we follow \citet{Indeb_05_extlaw} in
requiring signal-to-noise $\gtrsim$10 and paring the catalog to those stars with $\sigma$ $\lesssim$ 0.2 mag.  
To remove sources with possible intrinsic IR-excesses (e.g., YSOs or evolved stars with circumstellar shells), 
we adopt color restrictions of (\ione-\itwo) $\leq$ 0.6 mag and (\ithree-\ifour) $\leq$ 0.2 mag \citep{Flaherty_07_extlaw}.
Figure~\ref{fig:cmd}b shows the stars remaining after these photometric quality and color limits are imposed.

To ensure the purest possible sample of RC stars, we require that RC sample candidates have the characteristics of RC stars in both observed and intrinsic color space.
To that end, we deredden the remaining stars with a new technique, detailed elsewhere (Majewski et al., in prep), and then select only those stars
whose dereddened colors fall within the expected range of RC colors.  
In brief, this dereddening technique takes advantage of the fact that infrared photometry ($\lambda$ $\gtrsim$ 1.5 $\mu$m) measures primarily the 
Rayleigh-Jeans portion of stellar spectral energy distributions, which has a constant shape only minimally dependent on intrinsic stellar atmospheric variations; 
this constancy implies a homogeneous set of infrared colors, particularly at the longer wavelengths sampled by the {\it Spitzer}-IRAC bands.  
Thus, any departure from the nominal infrared colors measures the amount of reddening and extinction towards each star individually.  
We apply this technique to the sample of stars remaining after the photometric quality and color cuts imposed above, 
and we then retain only stars with 0.55 $\leq$ $(J-K_s)_0$ $\leq$ 0.9 mag (Figure~\ref{fig:cmd}c).
This step is invoked to help remove marginally-reddened RGB stars or highly-reddened bluer dwarfs that overlap the RC in the {\it observed} CMD.
We note that the potential circularity of selecting stars dereddened with an assumed extinction law is not a source of concern, because
the generous $\Delta(J-K_s)_0$ = 0.35 mag RC selection bin will still conservatively include very nearly all RC stars anyway.

Next, we divide the catalog into $\sim$2.5$^\circ$x2\degree blocks (in Galactic longitude and latitude, respectively) and plot a CMD for each block; 
straggler stars lying outside the obvious RC locus, which are primarily redder giant stars, 
as well as the brightest (unreddened) RC stars are then removed manually (Figure~\ref{fig:cmd}d).
Because many fainter stars unable to be positively associated with the RC locus are removed, this step has the effect of reducing the overall range of 
2MASS photometric uncertainties for stars ultimately used in the sample analysis.  Although high uncertainties in the PSC, particularly in the midplane, 
are potential indicators of a flux overestimation bias\footnote{http://www.ipac.caltech.edu/2mass/releases/allsky/doc/sec5\_3a.html}, 
Figure~\ref{fig:sigrange} shows that our final distribution of NIR photometric uncertainties is well below the limit where this bias is a concern.

\begin{figure}[htp]
 \begin{center}
 \includegraphics[width=0.5\textwidth]{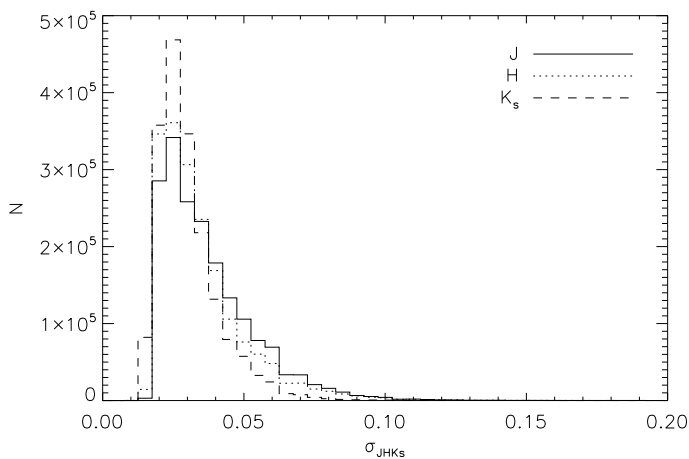}
 \end{center}
 \caption{Range of NIR photometric uncertainty in final RC stellar sample, for comparison to original uncertainty requirements ($\sigma_{JHKs}$ $\leq$ 0.4 mag).}
 \label{fig:sigrange}
\end{figure}

Finally, in order to make reliable color-color linear fits (Section~\ref{sec:CERfit}), 
we require an average RC reddening of at least 0.35 mag and a minimum {\it spread} in RC reddening of at least
0.15 mag (i.e., $\overline{(J-K_s)}$ $\geq$ 1 and $\Delta E(J-K_s)$ $\ge$ 0.15 mag) in each block's CMD.
We performed the color-color fits described below using a variety of photometric quality cuts, and the 
set of restrictions detailed here is the most lenient one (i.e., includes the most stars to provide better statistics)
yielding the same fit results as stricter requirements.

\subsection{Color Excess Ratios} \label{sec:CERfit}
A star's color excess $E(\lambda_1-\lambda_2)$ is simply the difference between the intrinsic color $(\lambda_1-\lambda_2)_0$
and the observed color $(\lambda_1-\lambda_2)$;
this excess depends on both the column density of dust along the line of sight to the star 
and the difference in extinction (for the same amount of dust) between $\lambda_1$ and $\lambda_2$.
However, the color excess {\it ratio} (CER) removes the total column density 
dependence (Appendix~\ref{sec:app:cer}) 
and reflects the differential behavior of extinction as a function of wavelength.
An example of a CER$_\lambda$ for a source is
\begin{equation} \label{equ:cer}
CER_\lambda=\frac{E(H-\lambda)}{E(H-K_s)}=\frac{(H-\lambda)-(H-\lambda)_0}{(H-K_s)-(H-K_s)_0},
\end{equation}
where $(H-\lambda)_0$ and $(H-K_s)_0$ are the intrinsic colors of the source.
Here we have chosen $E(H-K_s)$ to be the common reference denominator for our CER$_\lambda$ analysis because $(H-K_s)_0$ is a well-defined color for RC giants, 
more well-measured in our sample than MIR colors,
and less susceptible to color variations from metallicity, temperature, or non-RC contamination than the other NIR colors $(J-K_s)_0$ and $(J-H)_0$.  

Equation~\ref{equ:cer} can be rearranged as
\begin{equation}
(H-\lambda)=\frac{E(H-\lambda)}{E(H-K_s)}[(H-K_s)-(H-K_s)_o]+(H-\lambda)_o,
\end{equation}
where $(H-\lambda)$ and $(H-K_s)$ are the observed stellar colors.
For this work, we have adopted intrinsic colors for RC stars from the Padova suite of stellar models, where
$(H-K_s)_0$ $=$ 0.10 nearly independently of RC metallicity \citep[e.g.][]{Girardi_02_isochrones}.
When pairs of observed colors ($[H-K_s]$ and $[H-\lambda]$, for $\lambda$=J, \ione, \itwo, \ithree, \ifour) are plotted against one another for stars spanning a wide
range of extinctions, we find extremely linear correlations; a linear fit to these distributions gives estimates of the CER$_\lambda$ values as the slope, 
and the intrinsic $(H-\lambda)_0$ colors as the y-intercept when $(H-\lambda)$ is on the ordinate.  
Figure~\ref{fig:lfit} shows an example of these linear correlations and fits.

\begin{figure}[htpb]
 \begin{center}
 \includegraphics[angle=90,width=0.5\textwidth]{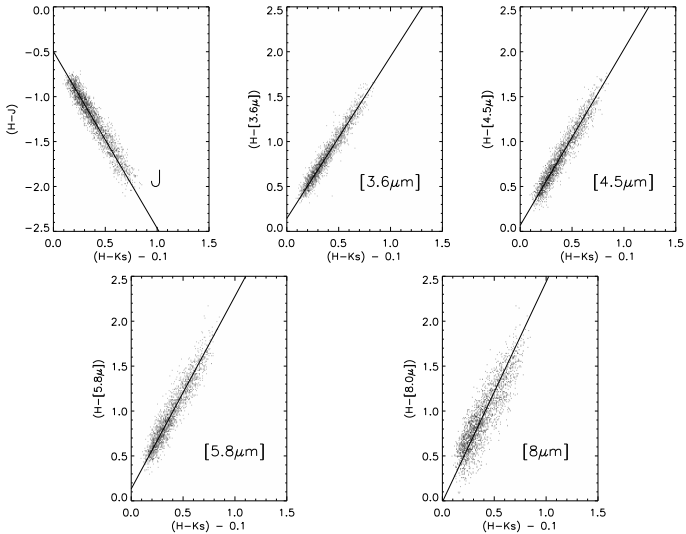}
 \end{center}
 \caption{Example of CER$_\lambda$ fits for a bin centered at $(l=30^\circ,b=0^\circ)$.  
The abscissa is $(H-K_s)-(H-K_s)_0$, where we have adopted $(H-K_s)_0$ $=$ 0.10 \citep[][]{Girardi_02_isochrones}.
The slope of each fit gives the color excess ratio for that bandpass, and the $y$-intercept gives the intrinsic $(H-\lambda)_0$ color for stars in this field.}
 \label{fig:lfit}
\end{figure}

Our goal is to measure CER$_\lambda$ values using RC stars in many different parts of the Galactic plane and to look for variations.
Before fitting, the data are binned by Galactic longitude into the same $\sim$2.5$^\circ\times$2\degree bins used in the RC sample selection.
By adopting only fields meeting the average-reddening and differential-reddening requirements 
(as before: $\overline{E(J-K_s)}$ $\ge$ 0.35 and $\Delta E(J-K_s)$ $\ge$ 0.15 mag), 
we fit 65 longitude bins, spanning from 10$^\circ$ to 100$^\circ$ from the Galactic center 
(i.e., 10 $\leq$ $l$ $\leq$ 65$^\circ$, $l$ = 90$^\circ$, and 260 $\leq$ $l$ $\leq$ 350$^\circ$).  
For each bin, we fit a set of five CER$_\lambda$ (i.e., [$H-K_s$] vs. each of [$H-\lambda$], $\lambda$ $\neq$ $K_s$)
using a linear weighted least-squares algorithm.
We take the uncertainties in the fits to be 1-$\sigma$, where $\sigma$ is the standard deviation of the fitted parameters.
Because of the intrinsic stellar density gradient in the disk, the number of stars in each bin varies (from 820 to $>$6$\times$10$^4$) across our longitude range.
We check for any effects this variation in sample size has on the fits by ``normalizing'' to the number of stars in the least-populated longitude bin (820);
that is, for every other bin, we apply the fitting algorithm to 820 randomly-selected stars a total of 25 times.
Though we find no noticeable difference with the parameters derived by fitting all the stars in each bin,
the results described below are the mean of the 25 fits for each bin, and the quoted uncertainties are either the average 1-$\sigma$
or the standard deviation of the 25 fit parameters, whichever is larger.

\subsection{Results} \label{sec:results_l}

\subsubsection{Derived CER$_\lambda$}

\begin{figure}[htpb]
 \begin{center}
 \includegraphics[width=0.5\textwidth]{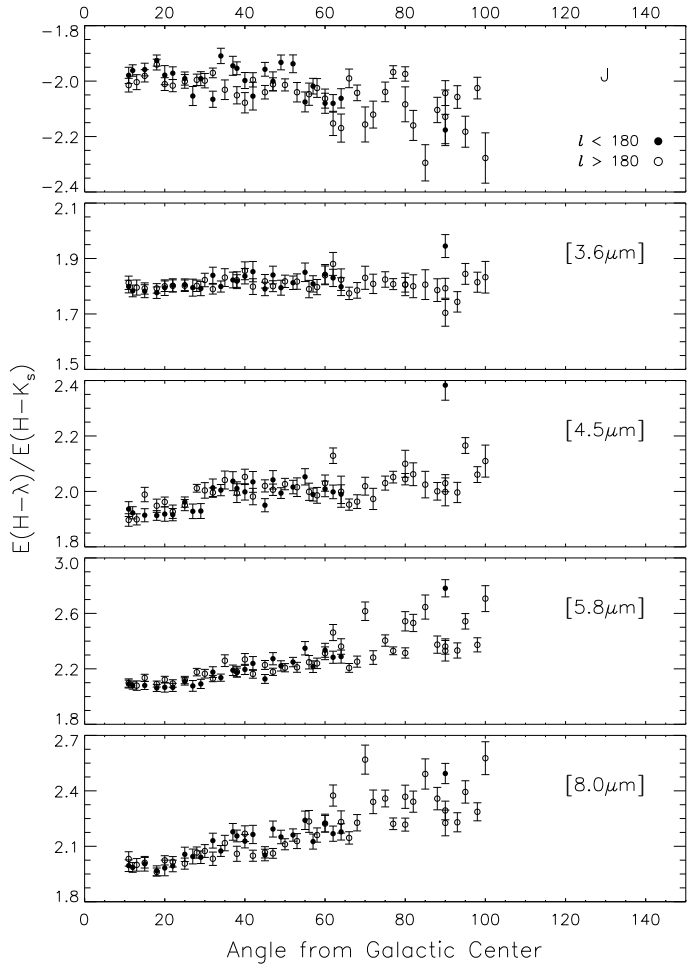}
 \end{center}
 \caption{CER$_\lambda$ $=$ $E(H-\lambda)$/$E(H-K_s)$ as a function of angle from the Galactic center (bins are 2.5$^\circ$ wide).  
The filled circles indicate data with $l$ $<$ 180 (Galactic quadrant I), and the open circles indicate data with $l$ $>$ 180 (Galactic quadrants III/IV).
There is no obvious difference in CER$_\lambda$ behavior between one side of the Galaxy and the other.
The error bars indicate either the 1-$\sigma$ uncertainty on the fit parameter or the standard deviation of the parameters from the 25 fits, whichever is larger.  
Notice the near constancy at 3.6\mic and the increasing steepness of the trend with increasing $\lambda$.}
 \label{fig:certrend}
\end{figure}

Figure~\ref{fig:certrend} shows the derived CER$_\lambda$ values as a function of angle from the Galactic center.  
This figure straightforwardly demonstrates the existence of a large-scale longitudinal variation in the infrared extinction law, symmetric about the Galactic center, 
and that the slopes of the CER$_\lambda$ trends increase as $\lambda$ increases (i.e., more drastic CER$_\lambda$ variations at longer MIR wavelengths).
Because an increased CER$_\lambda$ for $\lambda$ $>$ $H$ (and decreased CER$_\lambda$ for $\lambda$ $<$ $H$) results from relatively greater reddening  $E(H-\lambda)$,
these results indicate a steeper extinction curve ($A_\lambda$/$A_{Ks}$; Section~\ref{sec:extlaw}) in the outer Galaxy.
The CER$_\lambda$ values, averaged in 5$^\circ$ bins, are also presented in Table~\ref{tab:cer}.

\begin{deluxetable*}{c c c c c c}[htpb] \tablewidth{0pt}
\tablecaption{Derived CER$_\lambda$ values for use in Equation~\ref{eqn:cer2alak}.  For the contiguous data, the results have been averaged over 5$^\circ$.} 
\tablehead{\colhead{Galactic Longitude, $l$} & \colhead{CER$_J$} & \colhead{CER$_{[3.6\mu]}$} & \colhead{CER$_{[4.5\mu]}$} & \colhead{CER$_{[5.8\mu]}$} & \colhead{CER$_{[8\mu]}$}}
\startdata
10--15 & -1.97 & 1.79 & 1.92 & 2.08 & 2.00 \\
15--20 & -1.95 & 1.78 & 1.92 & 2.07 & 1.99 \\
20--25 & -1.98 & 1.80 & 1.93 & 2.08 & 2.01 \\
25--30 & -2.01 & 1.80 & 1.94 & 2.09 & 2.05 \\
30--35 & -1.99 & 1.82 & 2.01 & 2.16 & 2.10 \\
\hline
35--40 & -1.97 & 1.83 & 2.02 & 2.19 & 2.15 \\
40--45 & -2.00 & 1.83 & 1.99 & 2.19 & 2.11 \\
45--50 & -1.96 & 1.81 & 2.00 & 2.21 & 2.13 \\
50--55 & -2.01 & 1.83 & 2.03 & 2.30 & 2.20 \\
55--60 & -2.06 & 1.83 & 2.02 & 2.30 & 2.20 \\
\hline
60--65 & -2.07 & 1.82 & 2.00 & 2.30 & 2.19 \\
90 & -2.18 & 1.94 & 2.38 & 2.78 & 2.49 \\
260--265 & -2.16 & 1.83 & 2.11 & 2.54 & 2.42 \\
265--270 & -2.10 & 1.77 & 2.05 & 2.39 & 2.29 \\
270--275 & -2.14 & 1.77 & 2.01 & 2.43 & 2.34 \\
\hline
275--280 & -2.13 & 1.80 & 2.06 & 2.51 & 2.35 \\
280--285 & -2.02 & 1.81 & 2.06 & 2.40 & 2.29 \\
285--290 & -2.11 & 1.82 & 2.01 & 2.43 & 2.42 \\
290--295 & -2.06 & 1.80 & 1.98 & 2.36 & 2.31 \\
295--300 & -2.13 & 1.85 & 2.05 & 2.38 & 2.28 \\
\hline
300--305 & -2.05 & 1.81 & 2.00 & 2.27 & 2.21 \\
305--310 & -2.03 & 1.82 & 2.02 & 2.21 & 2.12 \\
310--315 & -2.02 & 1.81 & 2.02 & 2.20 & 2.08 \\
315--320 & -2.04 & 1.82 & 2.02 & 2.22 & 2.09 \\
320--325 & -2.05 & 1.84 & 2.03 & 2.24 & 2.12 \\
\hline
325--330 & -2.00 & 1.81 & 2.01 & 2.18 & 2.07 \\
330--335 & -2.00 & 1.81 & 1.99 & 2.15 & 2.05 \\
335--340 & -2.01 & 1.80 & 1.95 & 2.11 & 2.02 \\
340--345 & -1.98 & 1.79 & 1.97 & 2.12 & 2.00 \\
345--350 & -2.00 & 1.80 & 1.93 & 2.10 & 2.01
\enddata
\label{tab:cer}
\end{deluxetable*}

CER$_J$ appears nearly constant (with large scatter) until the angle $\sim$ 55$^\circ$, when it becomes increasingly negative, while the MIR CER$_\lambda$ values
have a somewhat flat distribution only out to $\sim$25$^\circ$.  Of all the IR colors under consideration, $(H-J)$ is the most sensitive to spectral variations
due to differences (spatially-dependent or random) in RC metallicity and temperature.  
Thus the behavior of CER$_J$ is consistent with the presence of global extinction law trends, 
but because it is also the most susceptible to unaccounted-for RC population variations, 
we do not evaluate the significance of the different Galactocentric angle of the trend inflection at this time.

\subsubsection{Checks on (H-$\lambda$)$_0$}

\begin{figure}[htpb]
 \begin{center}
 \includegraphics[width=0.5\textwidth]{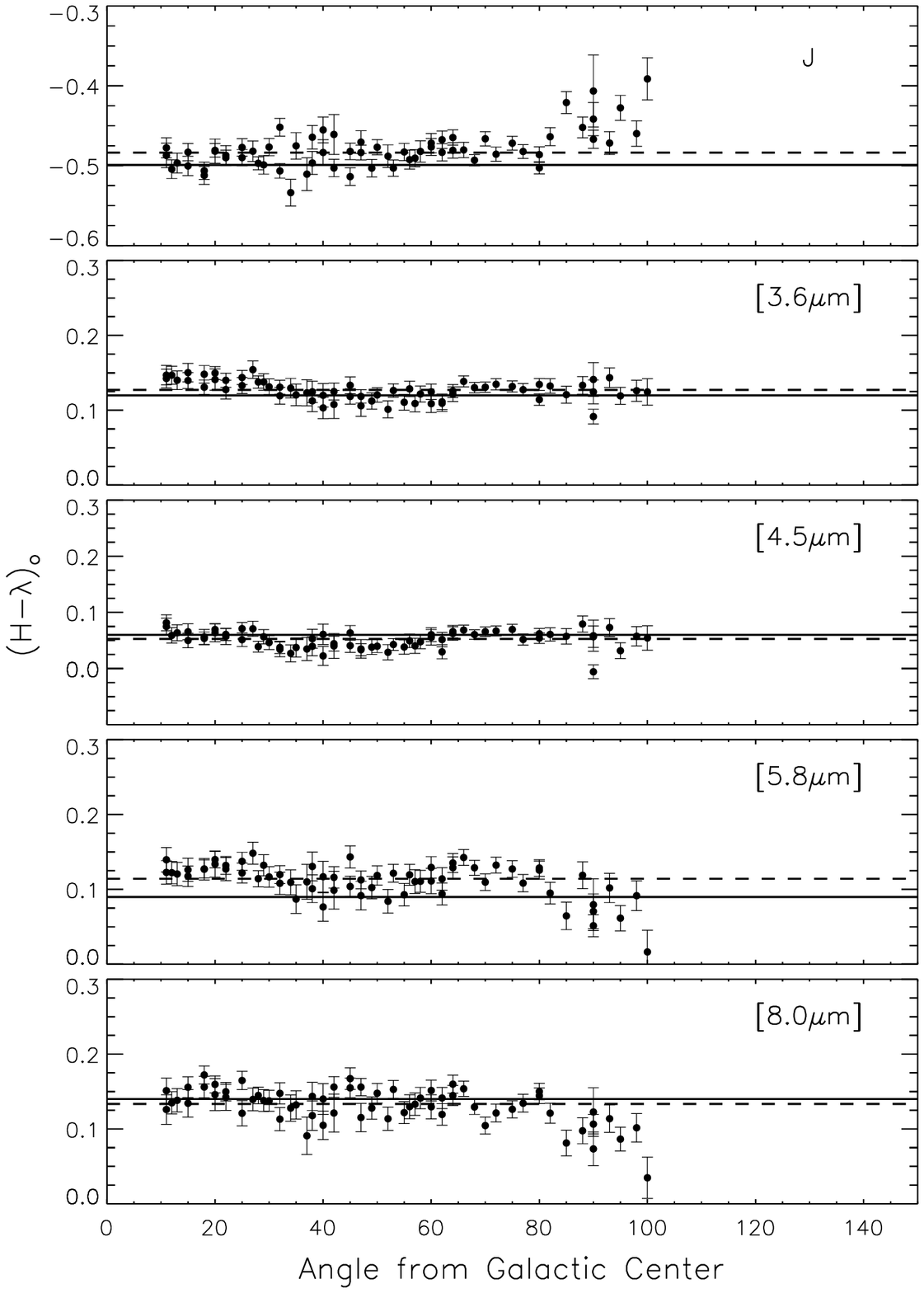}
 \end{center}
 \caption{The intrinsic colors of RC stars, measured as the y-intercepts in our color excess ratio fits, assuming a constant $(H-K_s)_0$=0.10 mag.
The error bars indicate either the 1-$\sigma$ uncertainty
on the fit parameter or the standard deviation of the parameters from the twenty-five fits, whichever is larger.
Also shown are the mean values for all longitudes (dashed lines) and the predicted intrinsic colors from the Padova stellar models
\citep[solid lines;][]{Girardi_02_isochrones}.}
 \label{fig:inttrend}
\end{figure}

In Figure~\ref{fig:inttrend}, we plot the derived intrinsic RC colors $(H-\lambda)_0$ against the field Galactocentric angle.  
Exploring the $y$-intercept values of our fits serves as a sanity check on our results---if our method is viable, then $(H-\lambda)_0$ from our fits 
should uniformly correspond to the intrinsic $(H-\lambda)_0$ colors of RC stars.
Figure~\ref{fig:inttrend} shows that we indeed find $(H-\lambda)_0$ colors that are not only consistent across our survey but also in
good agreement with the predictions of the Padova stellar models \citep[][]{Girardi_02_isochrones}.
Some slight differences ($\leq$0.025 mag) compared to the theoretical colors may be attributable to inconsistencies in 
bolometric corrections or inaccuracies in the filter transmission profiles applied to the stellar models.


It is possible, of course, that the true intrinsic colors of the RC stars in our sample do vary slightly, 
since the derived $(H-\lambda)_0$ values depend on our assumption of a constant $(H-K_s)_0$.  We validate the latter assumption using the recent
Milky Way disk abundance gradient of \citet{Pedicelli_09_d-feh} and the relationship between [Fe/H] and $(H-K_s)_0$ in the Padova isochrones.
For metallicities spanning -1.22 $\leq$ [Fe/H] $\leq$ +0.18, the intrinsic $(H-K_s)_0$ color of RC stars changes steadily by $\lesssim$0.06 mag.
\citet{Pedicelli_09_d-feh} find an [Fe/H] gradient of $\sim^-$0.13 dex/kpc inside the solar circle and $\sim^-$0.042 dex/kpc outside of it.  Given the approximate
range of $R_{GC}$ of our RC sample (see Section~\ref{sec:rgc}), this corresponds to a maximum $\Delta$[Fe/H] of 0.46 dex along any single line of sight, resulting in
a maximum intrinsic $(H-K_s)_0$ scatter of $\ll$0.06 mag in any single fitted field.  Since the measured $E(H-K_s)$ values are so much larger (usually $\geq$0.8 mag),
this possible scatter does not affect the derived CER$_\lambda$ values.  
This conclusion is further supported by the very tight linear color-color correlations observed (Figure~\ref{fig:lfit}),
indicating no significant intrinsic color variations within any single field.

\section{EXTINCTION AND DENSITY} \label{sec:density}

\subsection{Defining the Diffuse ISM} \label{sec:definediffuse}

Recent studies of dark molecular clouds have shown that the MIR extinction curve becomes increasingly shallow with 
increasing local dust density \citep[e.g.,][]{Flaherty_07_extlaw,RZ_07_extlaw,McClure_08_extlaw,Chapman_08_extlaw}.
Theoretical models predict this behavior as an effect of grain growth, from either the coagulation of grains or
the accumulation of refractory ice mantles in dense clouds \citep[][]{Li_03_dustoverview,Draine_03_dust}.  
In either case, the growth of grains serves both to remove small particles (including PAHs) and to add larger ones,
leading to an overall flatter (``grayer'') extinction.  In addition, the presence of certain ices may lead to enhanced extinction around their solid-state resonance bands 
(e.g., CO at 4.7 $\mu$m and H$_2$O at 3.1 and 6.0 $\mu$m).  However, neither the specific environmental conditions producing this grain growth, nor
the detailed behavior of $A_\lambda$/$A_{Ks}$ as a function of grain size, is yet well-constrained.  
\citet{Chapman_08_extlaw} approach these questions by using $A_V$ as a proxy for
density to probe specific molecular clouds and find a steady progression towards shallower 3.6-8\mic extinction curves as $<$$A_V$$>$ increases in the cloud.  
The 2.5$^\circ$-wide longitude bins and the large {\it total} spatial extent of our stellar
sample mean that any density variations potentially driving the extinction law variations seen here are likely dominated by 
the large-scale density gradient of the Galactic disk, not the small-scale filamentary structure of dark clouds.

This density gradient may be understood as either a change in the density of the mean ISM or a changing fractional contribution of molecular cloud material.
One of the stated goals of this study is to test the commonly used description of a simple, bimodal ISM comprising individually 
homogeneous ``dense'' regions (dark clouds and cores) and ``diffuse'' regions (everything in between).  
Thus, we must examine what we mean hereafter in this analysis when we refer to ``diffuse'' material and attempt to 
distinguish its reddening effects from those of denser clouds.  

Here, we explore the diffuse ISM using two separate methods to identify those stars tracing it.  Each of these---the ``RPD'' and ``$^{13}$CO'' methods---lead to their
own implicit definition of interstellar diffusivity, and we will discuss the relative merits of each method and definition.

\subsubsection{The RPD Method} \label{sec:rpd}

One approach to selecting stars reddened only by diffuse foreground material is to approximate the distance to each star 
and impose a maximum ``reddening per unit distance'' (RPD; e.g., $E[H-K_s]$/kpc) limit on the stellar sample; 
stars with RPD values greater than this limit would then be identified as being within or behind a ``dense cloud'' and removed from the analysis.
The major complication of this method 
is the uncertainty in the assumptions required to calculate the RPD limit for diffuse material.  

In the first place, one must know the distance to each star individually.  We estimate these values as described later in this paper (Section~\ref{sec:rgc}), 
but we need to note a critical source of systematic uncertainty.  Choosing an extinction law for each star is neither practical nor reliable, 
so a single law (extrapolated from $R_V$ = 3.1) is used for the entire RC sample.  
For lines of sight containing dense ISM (e.g., with $R_V$ $\sim$ 4-5), this choice has the end result of 
overestimating the stellar distances, thus underestimating the RPD values (by up to $\sim$30\%).  
Given the relatively narrow range of RPD values observed in our sample (0--0.5), this is a significant uncertainty.
So stars behind dense clouds are {\it systematically more likely} to 
be included in a diffuse ISM subsample defined by an upper RPD limit.

In addition, there is the process of even defining an appropriate RPD limit.  
One can either select characteristic values for the ISM with regard to number density ($n[H]$) and to $\frac{A_{Ks}}{N(H)}$ (or
the more common $\frac{A_V}{N(H)}$ and a conversion factor) or adopt a published RPD value.  
The extant literature contains ranges of values for all of these quantities such that the ``diffuse'' RPD 
may be said to span 0.05--0.22 mag kpc$^{-1}$ (a range of more than a factor of four), 
depending on results given by \citet{Cardelli_89_UV-O-IRextinction}, \citet{DickeyLockman_90_h1}, \citet{Madsen_05_diffuseISM}, \citet{Indeb_05_extlaw},
\citet{Misiriotis_06_ISMmodel}, or \citet{Pineda_08_COisotop}.

Given this wide range of realistic RPD values, and the open question of how the constituent factors may
vary with Galactic radius, the choice of a single RPD upper limit becomes tricky, even arbitrary.
Imposition of a particular limit has the potential to artificially skew the derived extinction law behavior 
by not accounting for density variations along the line of sight.  Moreover, making {\it a priori} assumptions
about the homogeneous properties of a likely heterogeneous ISM risks obscuring the very intrinsic characteristics of interstellar dust targeted by this study.

\subsubsection{The $^{13}$CO Method}

An alternative method of exploring the diffuse ISM is to identify (and remove) the effects of dense ISM on our extinction curves is 
through use of an appropriate tracer of high ISM density, such
as molecular gas emission.  The $J$=1$\rightarrow$0 transition of $^{13}$CO ($\nu_0$ = 110.2 GHz) is one such tracer, with an emission threshold corresponding to a cloud
density wherein there is only 1--2 mag of $A_V$ extinction.  
For a typical molecular cloud with $r$ $\sim$ 200 pc, this corresponds to a number density of roughly $n$(H) $\sim$ 1-5 cm$^{-3}$.
Comparison to the commonly used range of diffuse $n$(H) \citep[0.5--few cm$^{-3}$;][]{DickeyLockman_90_h1,Misiriotis_06_ISMmodel,Whittet_03_dust} demonstrates
$^{13}$CO to be a highly sensitive tracer of interstellar material dense enough to exhibit the known shallower extinction curves described above.
	
Given the ambiguities inherent in distances derived from molecular gas emission maps, we cannot conclusively match the $^{13}$CO emission 
to specific foreground dust clouds and remove only the stars behind them.
The most direct approach is then to remove all stars coincident in sky position with a $^{13}$CO detection.  
A disadvantage of this method is the exclusion of useful stars in intrinsically diffuse regions, in the foreground of denser material; 
nevertheless, the low optical depth and low emission
threshold of $^{13}$CO means that the resulting ``diffuse'' subsample is very conservatively defined and 
highly unlikely to be contaminated by stars in intrinsically dense regions.  
Another potential problem with using molecular gas as a tracer of dense ISM is the clumpy, filamentary structure of dark clouds and molecular cloud cores, 
which may prevent efficient filtering of dense ISM on a star-by-star basis.  
This obstacle, however, may be countered by using gas emission maps with resolution appropriate to trace molecular cloud structure at the median distance
of our stellar sample.

\subsection{Derived CER$_\lambda$ for Diffuse ISM} \label{sec:calcdiffuse}

To assess the effects of high-density clouds on our average ISM extinction curves, we sort our dataset into rough divisions of ``dense'' and ``diffuse'' lines of sight, 
separately for each of the two methods described above.
For the $^{13}$CO-emission method, we use the integrated-intensity maps of the Boston University-FCRAO Galactic Ring Survey 
\citep[GRS, spanning 18$^\circ$ $\leq$ $l$ $\leq$ 54$^\circ$;][]{Jackson_06_GRS} as tracers of high ISM density.
These maps have resolution high enough ($\sim$ 0.75') to distinguish filamentary 
structures of potentially dense ISM at the median distances of our stellar sample (where distances are estimated using the technique detailed in Section~\ref{sec:rgc}).  
As the most conservative selection of ``diffuse'' material, we consider only those stars with no measured $^{13}$CO emission
at their projected position on the sky (comprising $\sim$55\% of the RC stars in the longitudes covered by the GRS). 
As described above, this should eliminate any typically-sized molecular clouds with $n$(H) $\gtrsim$ 1-5 cm$^{-3}$ from affecting our derived extinction law.
For the RPD method, 
we consider two choices of maximum reddening-per-distance for ``diffuse'' ISM --- RPD $\leq$ 0.22 and RPD $\leq$ 0.15 mag kpc$^{-1}$ --- to 
assess the effects of the particular limiting value.
These cuts retain $\sim$79\% and $\sim$57\%, respectively, of the RC stars in the longitudes covered by the GRS.

We apply the CER$_\lambda$ fitting process as before (Section~\ref{sec:CERfit}) to these various ``diffuse'' subsamples.
The GRS coverage limits the $^{13}$CO-selection method to 18$^\circ$ $\leq$ $l$ $\leq$ 54$^\circ$, and we show results for the RPD method both with 
and without this longitude limitation.
The number of randomly selected stars for the 25 CER$_\lambda$ fits per field (i.e., the minimum starcount in any longitude bin) 
changes from 820 (Section~\ref{sec:CERfit}) to 14000 for RPD $\leq$ 0.22 mag kpc$^{-1}$, 
5700 for RPD $\leq$ 0.15 mag kpc$^{-1}$, and 8000 for $W$($^{13}$CO) $=$ 0 K; 
this increase in the number of fitted stars, compared to the original fits,
occurs because we are now considering only fields no further than 54$^\circ$ from the Galactic center.  
The increased counts are responsible for the smaller errorbars throughout Figure~\ref{fig:certrenddf} (compared to Figure~\ref{fig:certrend}).

\begin{figure*}[htpb]
 \begin{center}
 \includegraphics[width=0.6\textwidth, angle=90]{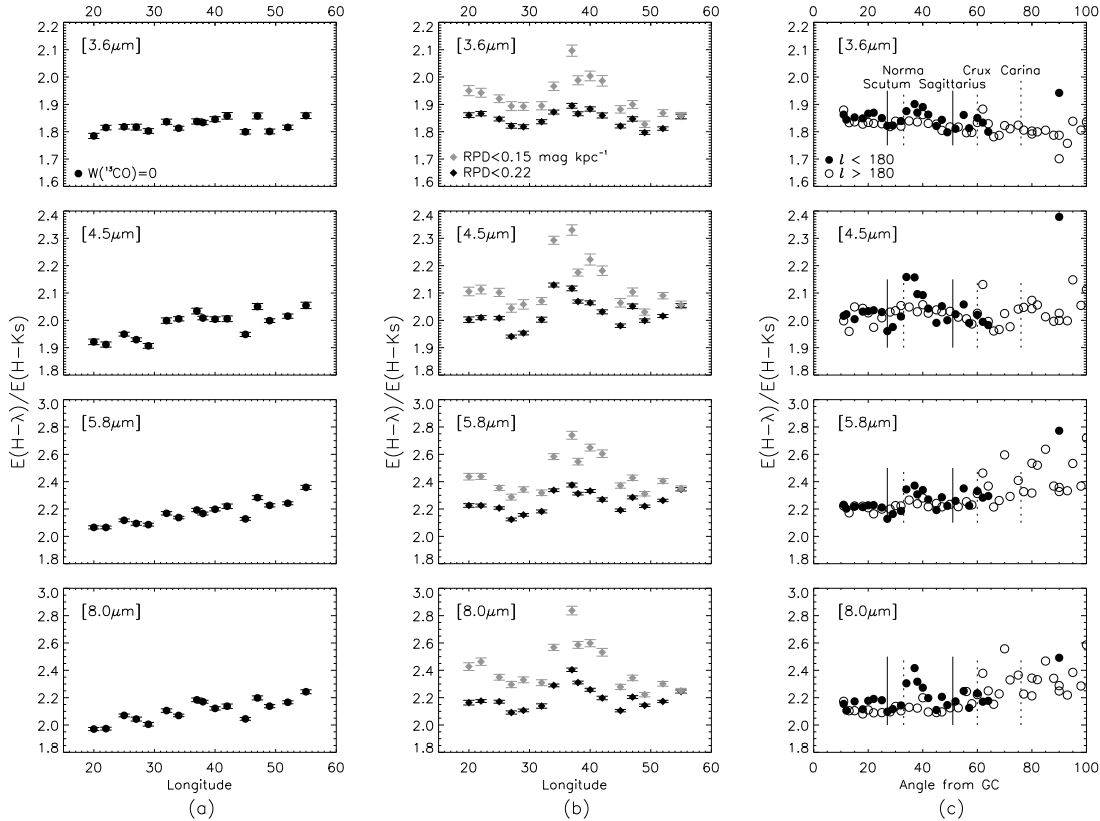}
 \end{center}
 \caption{CER$_\lambda$ $=$ $E(H-\lambda)$/$E(H-K_s)$ as a function of angle from the Galactic center for various definition of ``diffuse'' environments.
The error bars are as in Figure~\ref{fig:certrend}.  {\it Panel (a)}---Using stars selected as having no $^{13}$CO emission at their projected position on the sky.
{\it Panel (b)}---Using stars from the same sight-lines as in {\it (a)}, but with calculated reddening-per-distance (RPD) of $\leq$0.22 mag kpc$^{-1}$ (black diamonds) or $\leq$0.15 mag kpc$^{-1}$ (gray diamonds).  
{\it Panel (c)}---Using stars with RPD $\leq$ 0.22 mag kpc$^{-1}$ with $l$ $<$ 180$^\circ$ (filled circles) and $l$ $>$ 180$^\circ$ (open circles).  
The solid and dashed lines indicate the expected positions of spiral arm tangencies in Galactic quadrants I and IV, respectively.}
 \label{fig:certrenddf}
\end{figure*}

In the case where CER$_\lambda$ values are derived from stars with no $^{13}$CO emission detected at their positions (Figure~\ref{fig:certrenddf}a), we observe 
very nearly the same behavior with Galactic angle as seen in the case where no distinction between ``dense'' and ``diffuse'' lines of sight is made 
(i.e., Figure~\ref{fig:certrend}).
That is, even with the likeliest sites of larger dust grains explicitly removed, the ISM extinction behavior continues to depend on Galactocentric angle,
with the reddening curve becoming steeper toward the outer Galaxy.

For comparison, in the case where stars are filtered using the RPD method, 
the distribution of CER$_\lambda$ shows significant variation with longitude (Figure~\ref{fig:certrenddf}b), though no global trends are apparent. 
In addition, we find the counterintuitive result that the departure from canonical ``diffuse ISM'' values \citep[e.g.,][]{Indeb_05_extlaw} 
actually {\it increases} as the restriction on maximum RPD is lowered from 0.22 to 0.15 mag kpc$^{-1}$.
Moreover, the extinction curves that derive from these ``pure'' diffuse-only samples are difficult to explain physically, 
as they are steeper than even a simple power law or $R_V$ = 3.1 dust model (e.g., Section~\ref{sec:models}).  
At least part of the larger scatter in CER$_\lambda$ is due to the uncertainties 
inherent to fitting a straight line (as in Figure~\ref{fig:lfit}) to the increasingly incoherent color-color distributions that are obtained 
when the high-reddening stars are preferentially removed and the color-color plots become dominated by the photometric errors of the stars.  
If one considers the lower CER$_\lambda$ values near $l$ $\sim$ 28$^\circ$ and 50$^\circ$ to be troughs, 
rather than attributing the values at $l$ $\sim$ 37$^\circ$ to a peak, 
these depressions do correspond roughly with the expected positions of the tangencies of the Scutum and Sagittarius spiral arms \citep[e.g.,][]{Englmaier_99_arms}.
This might be one possible explanation for the troughs since decreased CER$_\lambda$ values indicate 
a shallower extinction curve (see Section~\ref{sec:extlaw}), consistent with what would be expected in a dense spiral arm.

However, ascribing this behavior to the spiral arms has many problems.  
First, because the removal of stars with high RPD estimates is explicitly intended to filter out all stars in or behind molecular clouds, it is unlikely that features 
caused by molecular clouds would become increasingly prominent as the RPD limit is lowered (in theory producing an even purer diffuse-only stellar sample).  
Second, while the precise heliocentric distances of these spiral arm tangent points are not tightly constrained, recent models place them near or 
beyond the range of our RC stars \citep[e.g.,][]{Hou_09_arms}, making it impossible for the stars with low RPD values in our dataset to probe spiral arm tangencies.  
Third, we can check for the corresponding spiral arm tangencies in the fourth Galactic quadrant 
(Carina, $l$ $\sim$ 284$^\circ$ and Crux, $l$ $\sim$ 310$^\circ$), since the RPD method is not limited to the longitude range imposed by the $^{13}$CO maps.  
Figure~\ref{fig:certrenddf}c shows CER$_\lambda$ values derived from stars 
with RPD $<$ 0.22 mag kpc$^{-1}$ over the entire longitude span of our data---no dips corresponding to the Crux and Carina arms are visible.  
In light of these inconsistencies,
we cannot ascribe the CER$_\lambda$ behavior for 18$^\circ$ $\leq$ $l$ $\leq$ 54$^\circ$ to any known Galactic structure.  The physical reality of the
observed features themselves (Figures~\ref{fig:certrenddf}bc) are called into question by 
(a) the number and type of assumptions that must be made to select a diffuse RPD limit from a wide (and potentially varying) range of possibilities, 
(b) the increasing departure from realistic CER$_\lambda$ values as the RPD limit is tightened, and (c) the strong dependence of the calculated RPD values on the stellar
distance estimates, which themselves require an assumed extinction law to calculate.

Given this wide range of complications, we conclude that the $^{13}$CO selection filter, while not perfect, has fewer systematic uncertainties than does the RPD filter,
and we therefore adopt the lack of $^{13}$CO emission as the preferred tracer of diffuse, intra-cloud interstellar material.
The similarity in behavior between this ``diffuse'' extinction and the ``all stars'' extinction derived in Section~\ref{sec:measure_ext} indicates that the environments
probed by our full set of RC stars are dominated by ISM too thin to harbor $^{13}$CO molecules;
the following analyses, therefore, are not limited to the available longitudes of the $^{13}$CO data 
and are able to explore the extinction behavior across the largest possible span of the Galactic disk.
The persistence of the trends between relative extinction and longitude in the diffuse ISM, particularly at 5.8
and 8 $\mu$m, indicates effects on the reddening law of either the mean Galactic density gradient (beyond the known growth effects in very dense clouds) 
or of secondary factors that vary throughout the Galactic disk, such as the chemical composition, size, or crystalline fraction of the dust grains.
These secondary factors, particularly grain size and crystallinity, 
may of course be at least partially due to the overall Galactic ISM density gradient (see the beginning of Section~\ref{sec:definediffuse}).  

For now, we adopt a working definition of ``diffuse ISM'' to refer to interstellar matter with a hydrogen number density too low to shield
$^{13}$CO molecules from dissociation (i.e., $n$(H) $\lesssim$ 1--5 cm$^{-3}$).
We believe that this definition is not only close to what is normally envisioned as ``diffuse ISM'' but also that which is most useful in
terms of practical and immediate applicability of our results to common dereddening problems.
In the 4-phase ISM model \citep[e.g.,][]{Whittet_03_dust}, material defined thusly corresponds to a mixture of the ``warm'' and ``cool (atomic)'' phases.  
We recognize that this thin matter is likely to have a large-scale radial density gradient throughout the disk, 
since it probably contains the outskirts of molecular clouds, which have a varying filling factor throughout the disk.
However, we emphasize that by our (or nearly any) definition this ISM would not be considered ``dense cloud'' material and so, prior to this study, 
would have been assigned a constant ``diffuse'' extinction law, regardless of its location in the Galaxy.
Our results here show that this constant extinction law does not accurately describe the extinction behavior along all lines of sight,
and any reddening corrections applied without considering the varying nature of the diffuse extinction law could be potentially subject to severe systematic errors.

\section{DISCUSSION} \label{sec:discuss}

\subsection{Conversion to $A_\lambda$/$A_{Ks}$} \label{sec:extlaw}

To convert the derived CER$_\lambda$ into the more familiar IR extinction law format $A_\lambda$/$A_{Ks}$, we use the relation
\begin{equation} \label{eqn:cer2alak}
\frac{A_\lambda}{A_{K_s}}=\frac{A_H}{A_{K_s}}-\left(\frac{A_H}{A_{K_s}}-1\right)\cdot CER_\lambda.
\end{equation}

Deriving the true, longitude-dependent extinction law requires an independent determination of $A_H$/$A_{Ks}$ along each line of sight for 
which we have measured CER$_\lambda$.  Unfortunately, this is an extremely challenging measurement to make.  \citet{Nishi_06_extlaw,Nishi_09_extlaw} have done it 
towards the Galactic center by assuming that all RC stars are at the same distance; we can make no such assumption for our lines of sight.
\citet{Indeb_05_extlaw} fit the RC locus in NIR CMDs to directly extract $A_H$/$A_{Ks}$, leaving distance as a free parameter and assuming that the amount of
extinction in those bands is constant per unit distance along the line of sight, the equivalent of assuming a smooth, homogeneous dust distribution.
This assumption is not generally applicable, and in our more extensive disk survey we see many lines of sight with CMDs containing shifts and kinks in the RC locus that
provide definitive evidence for a nonhomogeneous extinction and dust distribution.  
Thus, given no currently reliable method for determining $A_H$/$A_{Ks}$ around the disk, 
we cannot provide an absolute, independently calibrated set of $A_\lambda$/$A_{Ks}$ curves at this time.

Nevertheless, to facilitate comparison of $A_\lambda$/$A_{Ks}$ behavior with other studies, we use the commonly-adopted
value $A_H$/$A_{Ks}$ $=$ 1.55 as determined by \citet{Indeb_05_extlaw}, which corrresponds to an $A_\lambda\propto\lambda^{-\beta}$ power law with $\beta$ = 1.66.
The range of typical $\beta$ values (1.6--1.8) give $A_H$/$A_{Ks}$ $=$ 1.52--1.6; when applied to our observed CER$_\lambda$ with Equation~\ref{eqn:cer2alak}, this
range adds a $\sim$3-15\% uncertainty to the calculated $A_\lambda$/$A_{Ks}$ values.  This uncertainty is several times smaller than the percentage change in 
$A_\lambda$/$A_{Ks}$ with Galactic angle due to the intrinsic CER$_\lambda$ trends we observe here.
In addition, we note that the choice of $A_H$/$A_{Ks}$ only sets the absolute value scale for the relative extinction law;
it is not sensitive to variations in extinction law shape.  

\subsection{Comparison to Other Observational Studies}

In Figure~\ref{fig:alakcp}, we compare our mean $A_\lambda$/$A_{Ks}$ curve, derived using the total RC sample at all Galactocentric angles,
with the results of \citet{Lutz_96_galcenter}, 
\citet[][fit from their Equation 4]{Indeb_05_extlaw}, \citet{Jiang_06_7-15ext}, \citet{Flaherty_07_extlaw}, and \citet{Chapman_08_extlaw}.
We find very similar values and overall curve shape to the extinction laws found by many of these authors, but a closer inspection
reveals that our data yield slightly lower relative extinction values in the mid-infrared (i.e., a steeper MIR extinction curve).  This is entirely in accordance
with the difference in environments probed by these various studies: the shallowest curve results \citep{Lutz_96_galcenter,Jiang_06_7-15ext,Flaherty_07_extlaw}
are derived using stars in the Galactic Center and behind dense star-forming regions, where dust grains are expected to be significantly different than in the disk's
diffuse ISM (Section~\ref{sec:density}).  

\begin{figure}[htpb]
 \begin{center}
 \includegraphics[width=0.5\textwidth]{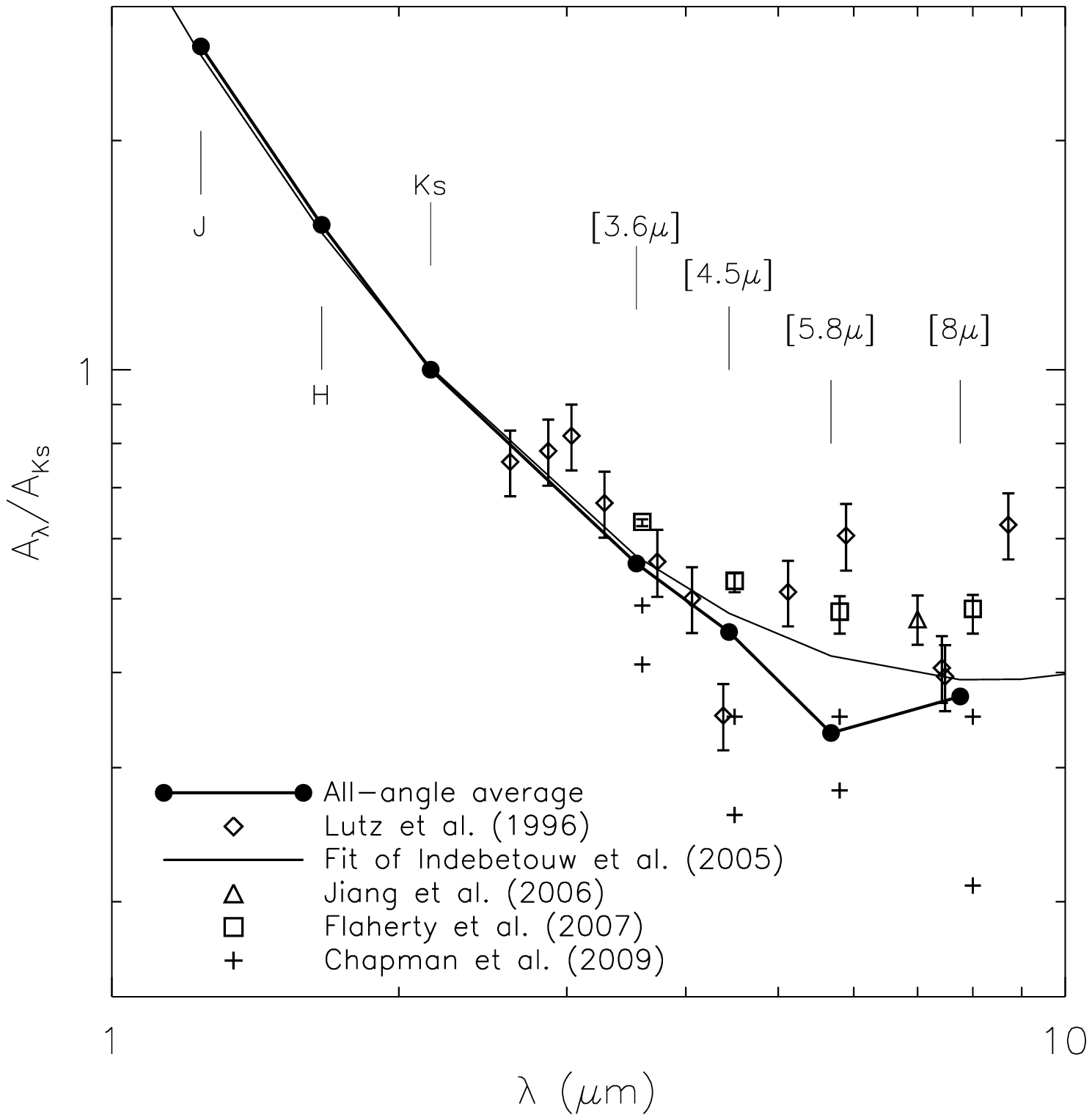}
 \end{center}
 \caption{Mean extinction values ($A_\lambda$/$A_{Ks}$) for this work (filled circles), compared to the Galactic center work of 
\citet[][diamonds]{Lutz_96_galcenter} and \citet[][triangles]{Jiang_06_7-15ext}, the ISM study of \citet[][thin line]{Indeb_05_extlaw}, and the dark cloud analyses of 
\citet[][squares]{Flaherty_07_extlaw} and \citet[][their $A_{Ks}$ $\leq$ 1 regions, pluses]{Chapman_08_extlaw}.  
The reported uncertainties in the \citet{Chapman_08_extlaw} data are large and have been excluded for clarity.}
 \label{fig:alakcp}
\end{figure}

Our results more closely match the curve of \citet{Indeb_05_extlaw}, which was derived using three lines of sight comprising
both diffuse ISM and an H{\small II} region, with Galactocentric angles $\sim$ 42$^\circ$ and 76$^\circ$.  
The fact that our mean curve lies below these data is entirely consistent with our relatively larger inclusion of diffuse and outer-Galaxy regions.
Another comparable match to our data comes from the two lowest-$A_{Ks}$ bins of \citet{Chapman_08_extlaw}, probing regions where $A_{Ks}$ $\leq$ 1.  The least-extinguished
stars in this dark-cloud survey are likely to be either foreground dwarfs unaffected by the molecular cloud material or stars observed through the low-density cloud edges.

\subsection{Comparison to Theoretical Models} \label{sec:models}

Because our study finds evidence for variations in the extinction law with longitude, it is useful to compare extinction curves from different sightlines
with curves extracted from theoretical dust models to attempt explanation of the observed differences.  
Figure~\ref{fig:alakcp2} shows such a comparison between the most current suite of dust 
models of \citet[][]{WeinDraine_01_dustsize}\footnote{http://www.astro.princeton.edu/$\sim$draine/dust/dust.html} and
our extinction curves sampled from various Galactocentric angles, ranging from the innermost GLIMPSE data (10$^\circ$--15$^\circ$ from Galactic center) 
to dust entirely beyond the solar circle ($>$90$^\circ$).
The Case A and Case B models of \citet{WeinDraine_01_dustsize} differ slightly in their carbon abundances (see below) and grain size 
restrictions---Case A contains grains only up to 1\mic while Case B grains can be as large as 10 $\mu$m.

\begin{figure}[htpb]
 \begin{center}
 \includegraphics[width=0.5\textwidth]{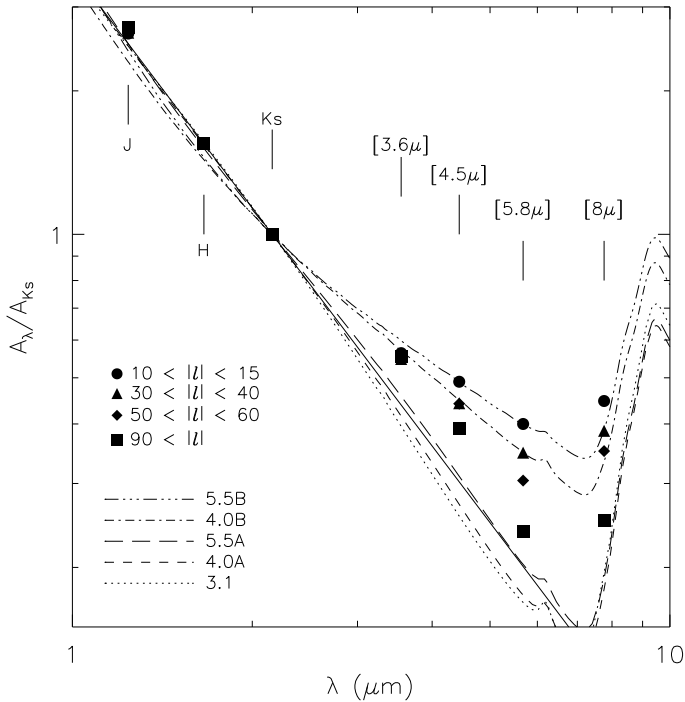}
 \end{center}
 \caption{Mean extinction curves for the indicated ranges of Galactic angle 
(smallest---circles; mid-range---triangles and diamonds; beyond the solar circle---squares),
compared to the current suite of theoretical dust models of \citet[][]{WeinDraine_01_dustsize}, with $R_V$ and size distribution case (A or B) as indicated.
Also shown for comparison is an $A_\lambda\propto\lambda^{-\beta}$ power law, for $\beta$ = 1.66 (solid line).}
 \label{fig:alakcp2}
\end{figure}

As in the case with the observational studies listed above, our results clearly show a shallower extinction law than either the $R_V$ = 3.1 theoretical model 
or any power law that could be extrapolated from the three NIR measurements alone (e.g., the solid line in Figure~\ref{fig:alakcp2}). 
The Case B models with $R_V$ = 4.0 and $R_V$ = 5.5 provide the closest overall matches to the ``inner'' Galactic sightlines, 
though these are nearly indistinguishable from each other bluewards of $\sim$3 $\mu$m;
we note that if we had scaled the theoretical curves using $A_H$/$A_{Ks}$ $=$ 1.55, as we did for our derived extinction curves, the MIR range of the Case B models
would shift downwards and align more closely with the $|l|$ $\sim$ 30--60$^\circ$ data points.
The original carbon abundances of the \citet{WeinDraine_01_dustsize} models has been called into question by observational constraints
\citep[see discussion in][]{Draine_03_dust}, 
but only those with $R_V$ = 3.1 or a Case A size distribution have subsequently been 
modified (with a change in C/H on the order of $\lesssim$15\%).
These Case A models more closely follow the $R_V$ = 3.1 curve, but all three yield MIR extinction curves too steep to explain our inner Galaxy observations.

Thus there is a clear relationship between Galactocentric angle and best-matched dust model: though the inner fields correspond to 
the higher $R_V$ values, the outer fields are increasingly similar to the highly-diffuse $R_V$ = 3.1 and the small-grain Case A models.  
The diffuse nature of outer disk Galactic dust is one of the reasons a large-scale extinction law survey of this type has not before been conducted---a diffuse ISM
produces small reddenings that are difficult to study, 
particularly in the relatively transparent MIR portion of the curve (compared to shorter, more dust-sensitive wavelengths);
our extensive disk coverage and high-quality MIR photometry allow us to make headway on overcoming these obstacles.
After determining extinction behavior in a consistent way across a wide swath of the Galactic disk and comparing the results to theoretical dust models, we 
interpret the variation in extinction law as a consequence of a decrease in mean dust grain size toward the outer Galaxy.

\subsection{8.0 $\mu$m Inflection}

\begin{figure}[htpb]
 \begin{center}
 \includegraphics[width=0.5\textwidth]{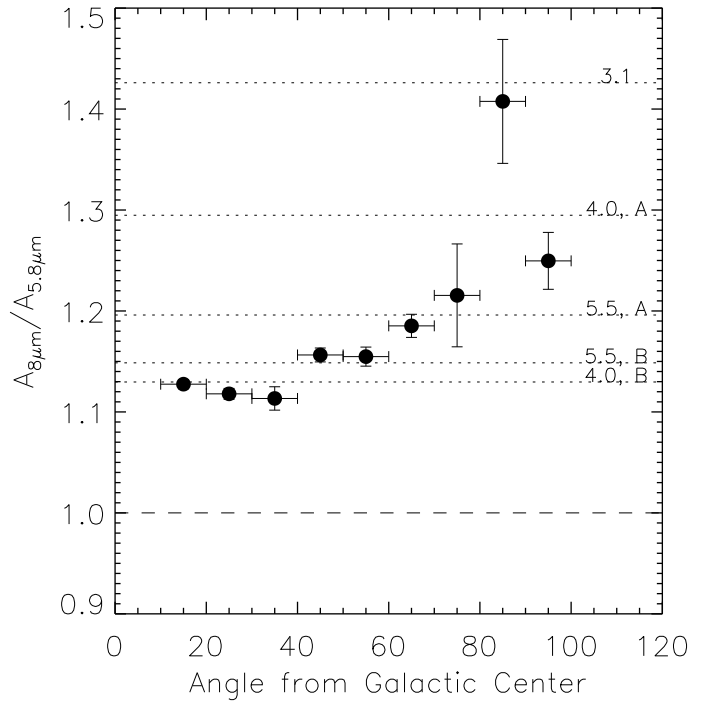}
 \end{center}
 \caption{Ratio between extinction at 8 and 5.8 $\mu$m.  The points are the mean values of the 2.5$^\circ$ fields in each 10$^\circ$ longitudinal bin, 
the horizontal bars indicate the degree span of each bin, and the vertical bars indicate the standard deviation of this ratio for the fields in each bin.  
The dashed line represents a truly ``flat'' extinction curve at these wavelengths, 
and the dotted lines represent approximate ratios for the theoretical dust models of \citet[][]{WeinDraine_01_dustsize}, with $R_V$ and size
distribution case as indicated.}
 \label{fig:8um}
\end{figure}

The presence of increased relative extinction in the IRAC-4 channel at $\sim$8\mic 
is generally attributed to the broad 9.7\mic resonance of amorphous silicates.  
\citet{Indeb_05_extlaw} estimate this absorption band has a $\lesssim$20\% effect on the expected
flux through the [8$\mu$] filter (but a negligible effect on the [5.8$\mu$] channel).  Though the exact behavior of the absorption band is loosely constrained
\citep[][]{Whittet_03_dust}, generally the band strength appears to decrease with both increasing grain size and increasing 
crystallinity \citep[Figure~\ref{fig:alakcp2}; models of][]{WeinDraine_01_dustsize}.  
Thus, one may expect a general trend of more dramatic 8\mic inflections with decreasing ISM density (loosely, with increasing Galactic angle),
and as shown in Figure~\ref{fig:8um}, we see just that.  
This figure shows the mean $A_{[8\mu]}$/$A_{[5.8\mu]}$ from our data for each 10$^\circ$ bin in Galactic angle,
and this ratio clearly increases towards the outer Galaxy.  Also included for comparison are the approximate corresponding ratios 
for the five dust models in Figure~\ref{fig:alakcp2} \citep{WeinDraine_01_dustsize}.  
These latter ratios are approximate because they do not take into account the true filter transmission profiles of the two IRAC channels; 
we use the FWHM and isophotal wavelength of each bandpass (convolved with a K2 giant star) to calculate the relative extinction values.
This analysis again suggests decreasing ISM grain size towards the outer Galaxy via the observed increasing 8\mic relative extinction and is further
evidence for the variable nature of the ISM constituents on Galactic scales.

\subsection{$A_\lambda$/$A_{Ks}$ as a Function of Galactocentric Radius} \label{sec:rgc}

Because each line of sight analyzed here thus far probes a range of Galactocentric radii ($R_{GC}$), the observed longitudinal variations in CER$_\lambda$
demonstrate that the assumption of a universal infrared extinction law, even for diffuse material, is incorrect---if it were otherwise, 
then the same extinction curve would be measured in all directions in the Galactic plane and at all distances.
Given that RC stars are good standard candles and that the factors affecting $A_\lambda$/$A_{Ks}$ 
(e.g., dust composition, grain size) are likely to have a radial dependence, 
the possibility of measuring the IR extinction law as a function of Galactocentric radius is a tantalizing one.
The difficulty, of course, lies in the fact that the reddening we observe towards each star is caused by dust spread along the line of sight to the star, so that the
observed extinction behavior cannot be assigned to the dust at one particular $R_{GC}$.  

On the other hand, for certain lines of sight, we can make simple assumptions regarding the dust distribution
that allow reasonable estimates of trends in $A_\lambda$/$A_{Ks}$ with $R_{GC}$.
First, we assume that the large-scale Galactic dust distribution has a strong radial density gradient 
\citep[scalelength $\sim$ 2--3 kpc;][]{Drimmel_01_3DMW,L-C_02_MWscales,Misiriotis_06_ISMmodel} from the Galactic center to the outer disk.
In this case, for sightlines towards the Galactic center ($|l|$ $\leq$ 13$^\circ$), we can assign the majority of the dust to the approximate $R_{GC}$ of the farthest
concentration of stars in our sample, where the dust is presumably thickest (for our RC sample, $R_{GC}$ $\sim$ 5.5 kpc, assuming $R_{GC,\sun}$ = 8 kpc).  
Likewise, for large-angle sightlines ($\geq$150$^\circ$, including the Galactic anti-center), 
we estimate most of the dust to lie in the foreground of the survey stars nearest the sun (for our sample, $R_{GC}$ $\sim$ 10.4 kpc),
where we expect the dust density to be highest.
Note that we discard the minimum reddening requirements (\S\ref{sec:sample}) in order to use these outer Galaxy fields; 
the small stellar reddenings at these longitudes are responsible for the increased uncertainties in the linear color-color fit parameters.
Finally, there exists an optimal Galactic longitude at which the majority of the observed stars, at varying {\it heliocentric} distances, 
fall along the same {\it Galactocentric} ring (with a constant, well-defined $R_{GC}$).  
To calculate this, we recognize that each line of sight (with $|l|$ $\leq$ 90$^\circ$) intersects the tangent point of a Galactocentric ring, 
where a concentration of stars (and the intervening dust) have the same $R_{GC}$; 
generally there will also be stars and dust 
with larger $R_{GC}$ on the near and far sides of the tangent point (as seen from the sun's position).
But as Figure~\ref{fig:rgc_lon} demonstrates, there are longitudes at which other factors serve to remove stars in front of and behind the tangent point---at
these longitudes, nearer stars are either too bright for 2MASS/{\it Spitzer} or are blue main sequence stars, 
and farther stars are actually extinguished out of our database altogether.
For our particular sample's heliocentric distance distribution, 
this critical angle is 58$^\circ$ (i.e., $l$ = 58$^\circ$ and 302$^\circ$), 
corresponding to the ring at $R_{GC}$ $\sim$ 6.8 kpc and indicated by the dotted line in Figure~\ref{fig:rgc_lon}.

\begin{figure}[htpb]
 \begin{center}
 \includegraphics[width=0.5\textwidth]{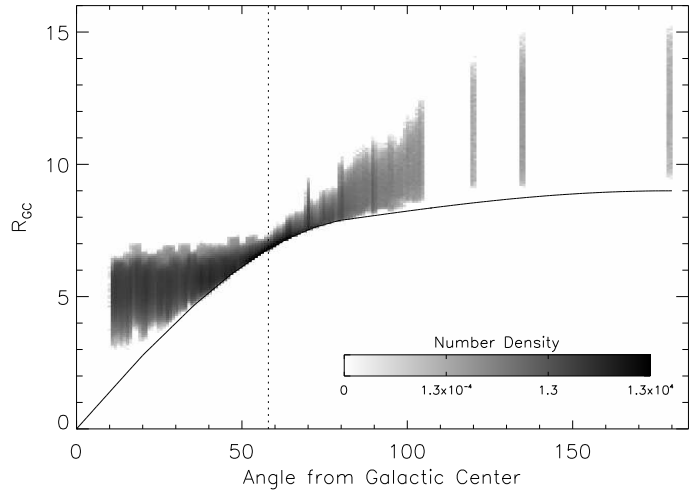}
 \end{center}
 \caption{The distribution of approximate Galactocentric radii ($R_{GC}$) for our RC sample, 
as a function of angle from the Galactic Center (\S\ref{sec:rgc}; assuming $R_{GC,\sun}$ = 8 kpc).  
The dotted line indicates the longitude ($|l|$ $\sim$ 58$^\circ$) at which the spread in $R_{GC}$ is smallest---that is, 
at which most stars along the line of sight are at a consistent, well-defined $R_{GC}$.  
The solid line indicates the limit of radii geometrically excluded from the sun's position.}
 \label{fig:rgc_lon}
\end{figure}

Distances to the RC stars are estimated using the extinction-measurement technique mentioned in Section~\ref{sec:sample} (to be detailed in Majewski et al., in prep);
the law of cosines is applied to convert the midplane heliocentric distances ($d$) into $R_{GC}$: 
\begin{equation}
R_{GC} = \sqrt{d^2+R_{GC,\sun}^2-2dR_{GC,\sun}\cos(l)}, 
\end{equation}
where $l$ is a star's Galactic longitude and $R_{GC,\sun}$ is assumed to be 8 kpc.
In converting reddenings to extinction, we must assume an absolute magnitude $M_{Ks}$ for red clump stars 
\citep[$M_{Ks}$ $\sim$ $-$1.54;][]{Groenewegen_08_RCmag}
and a NIR+MIR extinction law, for which we choose that of \citet{Indeb_05_extlaw} for all three lines of sight described above.  
We realize that this introduces a significant systematic uncertainty into our calculated $R_{GC}$ values, 
which varies depending on the relative molecular cloud filling factor along the line of sight (Section~\ref{sec:rpd});
however, we note that in choosing a consistent law for all three cases, 
the systematic differences in CER$_\lambda$ among them are still highly informative.
The approximate $R_{GC}$ values themselves are sufficient for our illustrative goals here, as we are not attempting to ``fit'' a global relationship but rather
to show relative trends.

Figure~\ref{fig:bydist2} shows the fitted mid-infrared CER$_\lambda$ values (i.e., no $A_H$/$A_{Ks}$ assumed) for the three ranges of Galactic angle described above.  
The doubled points for the inner angles correspond to the sightlines on either side of the Galactic center.  We clearly see a trend with Galactic radius, 
evidence for an intrinsic $A_\lambda$/$A_{Ks}$ radial dependence underlying the apparent dependence of the infrared extinction law on Galactic angle.  
The behavior with $R_{GC}$ is as expected from both the analyses in this section and theoretical
predictions: the presumably more diffuse ISM in the outer disk produces higher CER$_\lambda$ values 
(which appear as a steeper extinction curve) relative to the inner disk.
Comparison to theoretical models suggests consistency with the presence of larger grains and a 
higher selective-to-relative extinction ratio in the inner Galaxy, decreasing to almost exclusively sub-micron particles in the outer disk.

\begin{figure}[htpb]
 \begin{center}
 \includegraphics[width=0.5\textwidth]{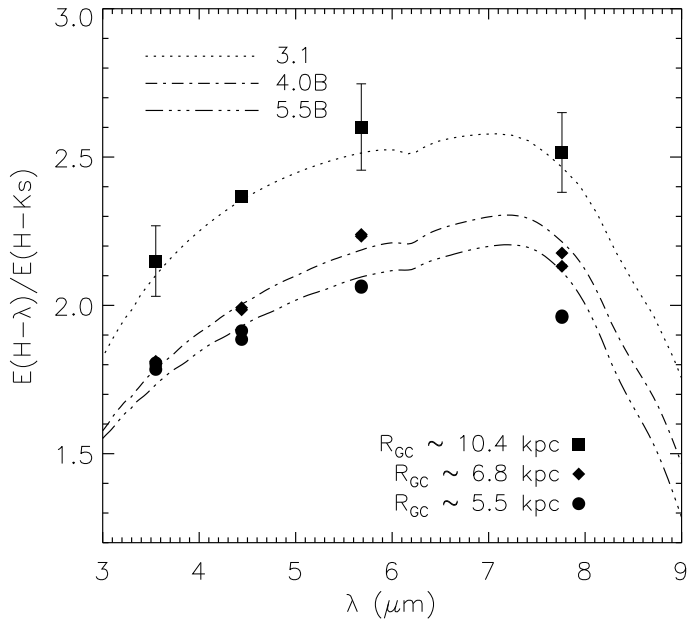}
 \end{center}
 \caption{CER$_\lambda$ values for the three ranges of Galactocentric angle ($\geq$150$^\circ$, 57$^\circ$--59$^\circ$, and $\leq$13$^\circ$, shown by squares, diamonds, and
circles, respectively) dominated by dust at specific radii $R_{GC}$.  
The double points for $R_{GC}$ $=$ 5.5 kpc and $R_{GC}$ $=$ 6.8 kpc correspond to the sightlines on either side of the Galactic center.  
Also shown are three curves extrapolated from the dust models by \citet[][]{WeinDraine_01_dustsize}, with $R_V$ and size distribution case as indicated.}
 \label{fig:bydist2}
\end{figure}

Interstellar dust grain size is a function of both the initial size distribution at formation and the ongoing processes that further affect grain size, crystallinity, and
even composition.
The formation size distribution of Galactic dust, and indeed the formation mechanism(s) itself, is not yet definitively understood
\citep[e.g.,][]{Speck_09_dustformCstars,Zhu_09_dustevolve}, and layered on top of these are multiple complex processes, including
grain coagulation, gas accretion, thermal annealing, collisional shattering, and gas-grain sputtering, which depend on local ISM phase and conditions.  
Our initial conclusions from the various analyses presented here regarding the mean grain size \emph{spatial} 
distribution---larger (at least micron-sized) grains at small Galactocentric radii, decreasing steadily to sub-micron ones past the solar circle---provide 
an observational constraint on the convolved patterns of grain size formation distribution 
and the relative importance of local conditions and processing mechanisms at different locations in the disk.

\section{SUMMARY AND CONCLUSIONS} \label{sec:conc}

Using data from 2MASS \citep{Skrutskie_06_2mass} and three extensive {\it Spitzer}-IRAC surveys \citep[][PID 20499, and PID 40791]{Benjamin_03_glimpse}, 
we have studied the interstellar relative extinction as a function of wavelength through the 
near- and mid-infrared (1.2-8 $\mu$m) for contiguous lines of sight covering $\sim$150$^\circ$ of the Galactic disk.
Using G and K spectral type red clump (RC) stars, 
we have calculated color excess ratios and derived the wavelength-dependent extinction behavior as a function of angle from the Galactic center.
We find the IR extinction law becomes increasingly steep as the Galactocentric angle increases, 
with identical behavior between $l$ $<$ 180$^\circ$ and $l$ $>$ 180$^\circ$.

Aware that dense molecular clouds scattered throughout the disk have recently been shown to 
exhibit MIR extinction behavior quite different from that of the diffuse ISM \citep[e.g.,][]{Flaherty_07_extlaw,RZ_07_extlaw,McClure_08_extlaw,Chapman_08_extlaw}, 
we culled from our sample those stars spatially associated with lines of sight having detectable $^{13}$CO ($J$=1$\rightarrow$0) emission \citep[][]{Jackson_06_GRS},
here used as an indicator of high ISM density ($n$(H) $\gtrsim$ 1--5 cm$^{-3}$).  After doing so, we find that
the derived $A_\lambda$/$A_{Ks}$ trend with Galactic angle persists, in what should be diffuse ISM only.  
This behavior suggests that, apart from high density molecular clouds, at least one secondary factor that varies 
throughout the Galactic disk is at work, such as grain size, composition, or crystalline fraction.
As these factors themselves depend on local ISM conditions (including density), 
and that even within the diffuse ISM there are likely to be density gradients,
we see clearly that the previous simple, bimodal division of the ISM into ``dense'' and ``diffuse'' environments
is not adequate to characterize the displayed variation in extinction behavior.  

We find close matches to our empirical extinction curves (as derived from our diffuse-only and total RC samples) in the diffuse-ISM study of \citet{Indeb_05_extlaw} 
and the lowest-density cases of \citet{Chapman_08_extlaw}, but not in the work of \citet{Lutz_96_galcenter}, \citet{Jiang_06_7-15ext}, and
\citet{Flaherty_07_extlaw}, who focus on regions of known high dust density.
Our results are consistent with several of the theoretical dust models presented by \citet{WeinDraine_01_dustsize}, 
and we use this comparison to characterize the dust at different positions in the Galaxy.
We find that the best-matched $R_V$ decreases (from 5.5 to 3.1) at increasingly greater angles from the Galactic Center, and the input size distributions of the best-fit
models suggest a decrease in mean grain size (from super- to sub-micron) towards the outer Galaxy.

We characterize the extinction ``upturn'' at $\sim$8 $\mu$m in each field's extinction law with the ratio $A_{[5.8\mu]}$/$A_{[8\mu]}$,
and we find a clear increase in the strength of this inflection at greater Galactocentric angles.
The 9.7\mic silicate absorption, to which we attribute this increased relative extinction, is strengthened by certain changes in dust grain 
characteristics (e.g., to a smaller mean size and/or lesser crystalline fraction); because these changes are consistent with the trends between Galactic longitude
and dust grain property derived from the model comparisons above, 
we interpret the 8\mic extinction inflection behavior as further evidence for the varying nature of ISM dust grains from the inner to the outer Galaxy

We consider the problem of calculating extinction curves as a function of true Galactic radius $R_{GC}$.  
This is the ideal, more natural approach to parameterizing the factors that cause the observed variations in extinction behavior, 
but assigning the dust spread out along the line of sight to a particular $R_{GC}$ is difficult to do.
To make a first attempt, we select three lines of sight for which we can make good approximations 
of the dust's typical $R_{GC}$ (at $\sim$5.5, $\sim$6.8, and $\sim$10.4 kpc), 
and we find that the trend in extinction behavior strongly supports a decrease in mean dust grain size (and in $R_V$) at greater Galactic radii.

Careful measurements of the extinction in the outer part of the Galaxy are difficult to make because of 
the smaller numbers of observed stars and the weaker extinction to those stars.  
Nevertheless, large-scale surveys of the outer disk, such as the approved GLIMPSE-360 Cycle 6 {\it Spitzer}-IRAC survey, 
will provide stellar photometry (at \ione and \itwo) of the most distant, reddened stars to the edge of the disk.  
A better understanding of the Milky Way extinction law as a function of Galactic position 
is crucial not only to a complete description of the varying dust grain characteristics but also to photometric and spectroscopic corrections made for observations 
of sources within or beyond our Galaxy that are affected by dust reddening and extinction.

\begin{acknowledgements}
We are grateful to P.~Arras for useful discussions, to J.~K.~Carlberg for comments on the manuscript, and to the anonymous referee for help with
improvements and clarifications within the paper.
We also thank the Spitzer Science Center for hosting GZ as a guest and SRM as a Distinguished Visiting Scientist in June 2008, 
during which period some of this research was conducted.
This work is based in part on observations made with the {\it Spitzer} Space Telescope, and has made use of the NASA~/~IPAC Infrared Science Archive, 
which are operated by the Jet Propulsion Laboratory, California Institute of Technology under a contract with NASA. 
Support for this work was provided by NASA through awards 1276756 and 1316912 issued by JPL/Caltech.
We acknowledge use of data products from the Two Micron All Sky Survey, a joint project of the University of Massachusetts and the 
Infrared Processing and Analysis Center~/~California Institute of Technology, funded by NASA and the NSF.  
This publication also makes use of molecular line data from the Boston University-FCRAO Galactic Ring Survey (GRS). 
The GRS is a joint project of Boston University and Five College Radio Astronomy Observatory, 
funded by the National Science Foundation under grants AST-9800334, AST-0098562, \& AST-0100793.  
GZ acknowledges support from the Virginia Space Grant Consortium.
\end{acknowledgements}

\begin{appendix}
\section{Column Density Independence of Color Excess Ratios} \label{sec:app:cer}
The extinction $A_\lambda$ at wavelength $\lambda$ is the difference between the source's intrinsic intensity $I_{0,\lambda}$ and the observed, extinguished intensity
$I_\lambda$=$I_{0,\lambda}e^{-\tau_\lambda}$:
\begin{eqnarray}
A_\lambda & = & -2.5\log{\frac{I_\lambda}{I_{0,\lambda}}} \nonumber \\
 & = & -2.5\log{e^{-\tau_\lambda}} \nonumber \\
\therefore A_\lambda & \sim & 1.086\tau_\lambda. 
\end{eqnarray}

For a given color $(\lambda_1-\lambda_2)$, the reddening $E(\lambda_1-\lambda_2)$ is the difference in extinction (in magnitudes) between $\lambda_1$ and $\lambda_2$:
\begin{equation}
E(\lambda_1-\lambda_2) = A_{\lambda_1}-A_{\lambda_2} = 1.086(\tau_{\lambda_1}-\tau_{\lambda_2})
\end{equation}

Recalling that $\tau_\lambda$ = $N\cdot\kappa_\lambda$, where $N$ (cm$^2$) is the column density of extinguishing material along the line of sight 
and $\kappa_\lambda$ (cm$^{-2}$) is the extinction cross-section, the color excess ratio (CER$_\lambda$) can be expressed as
\begin{eqnarray}
CER_\lambda & = & \frac{E(H-\lambda)}{E(H-K_s)} \nonumber \\
 & = & \frac{\tau_H-\tau_\lambda}{\tau_H-\tau_{Ks}} \nonumber \\
 & = & \frac{N\;(\kappa_H-\kappa_\lambda)}{N\;(\kappa_H-\kappa_{Ks})} \nonumber \\
CER_\lambda & = & \frac{\kappa_H-\kappa_\lambda}{\kappa_H-\kappa_{Ks}};
\end{eqnarray}
i.e., CER$_\lambda$ is independent of the line of sight's particular column density $N$.  
The CER$_\lambda$ is a function of the wavelength-dependent extinction cross-section per particle, which is governed by the grain composition, size, and shape, 
but not the total column density (or for this work, the potentially vastly different column densities towards each star in the sample).
\end{appendix}

\bibliographystyle{aa}

\begin{thebibliography}{35}
\expandafter\ifx\csname natexlab\endcsname\relax\def\natexlab#1{#1}\fi

\bibitem[{{Benjamin} {et~al.}(2003){Benjamin}, {Churchwell}, {Babler}, {Bania},
  {Clemens}, {Cohen}, {Dickey}, {Indebetouw}, {Jackson}, {Kobulnicky},
  {Lazarian}, {Marston}, {Mathis}, {Meade}, {Seager}, {Stolovy}, {Watson},
  {Whitney}, {Wolff}, \& {Wolfire}}]{Benjamin_03_glimpse}
{Benjamin}, R.~A., {Churchwell}, E., {Babler}, B.~L., {et~al.} 2003, \pasp,
  115, 953

\bibitem[{{Cardelli} {et~al.}(1989){Cardelli}, {Clayton}, \&
  {Mathis}}]{Cardelli_89_UV-O-IRextinction}
{Cardelli}, J.~A., {Clayton}, G.~C., \& {Mathis}, J.~S. 1989, \apj, 345, 245

\bibitem[{{Chapman} {et~al.}(2009){Chapman}, {Mundy}, {Lai}, \&
  {Evans}}]{Chapman_08_extlaw}
{Chapman}, N.~L., {Mundy}, L.~G., {Lai}, S.-P., \& {Evans}, N.~J. 2009, \apj,
  690, 496

\bibitem[{{Dickey} \& {Lockman}(1990)}]{DickeyLockman_90_h1}
{Dickey}, J.~M. \& {Lockman}, F.~J. 1990, \araa, 28, 215

\bibitem[{{Draine}(2003)}]{Draine_03_dust}
{Draine}, B.~T. 2003, \araa, 41, 241

\bibitem[{{Drimmel} \& {Spergel}(2001)}]{Drimmel_01_3DMW}
{Drimmel}, R. \& {Spergel}, D.~N. 2001, \apj, 556, 181

\bibitem[{{Englmaier} \& {Gerhard}(1999)}]{Englmaier_99_arms}
{Englmaier}, P. \& {Gerhard}, O. 1999, \mnras, 304, 512

\bibitem[{{Flaherty} {et~al.}(2007){Flaherty}, {Pipher}, {Megeath}, {Winston},
  {Gutermuth}, {Muzerolle}, {Allen}, \& {Fazio}}]{Flaherty_07_extlaw}
{Flaherty}, K.~M., {Pipher}, J.~L., {Megeath}, S.~T., {et~al.} 2007, \apj, 663,
  1069

\bibitem[{{Girardi} {et~al.}(2002){Girardi}, {Bertelli}, {Bressan}, {Chiosi},
  {Groenewegen}, {Marigo}, {Salasnich}, \& {Weiss}}]{Girardi_02_isochrones}
{Girardi}, L., {Bertelli}, G., {Bressan}, A., {et~al.} 2002, \aap, 391, 195

\bibitem[{{Groenewegen}(2008)}]{Groenewegen_08_RCmag}
{Groenewegen}, M.~A.~T. 2008, \aap, 488, 935

\bibitem[{{Hou} {et~al.}(2009){Hou}, {Han}, \& {Shi}}]{Hou_09_arms}
{Hou}, L.~G., {Han}, J.~L., \& {Shi}, W.~B. 2009, \aap, 499, 473

\bibitem[{{Indebetouw} {et~al.}(2005){Indebetouw}, {Mathis}, {Babler}, {Meade},
  {Watson}, {Whitney}, {Wolff}, {Wolfire}, {Cohen}, {Bania}, {Benjamin},
  {Clemens}, {Dickey}, {Jackson}, {Kobulnicky}, {Marston}, {Mercer},
  {Stauffer}, {Stolovy}, \& {Churchwell}}]{Indeb_05_extlaw}
{Indebetouw}, R., {Mathis}, J.~S., {Babler}, B.~L., {et~al.} 2005, \apj, 619,
  931

\bibitem[{{Jackson} {et~al.}(2006){Jackson}, {Rathborne}, {Shah}, {Simon},
  {Bania}, {Clemens}, {Chambers}, {Johnson}, {Dormody}, {Lavoie}, \&
  {Heyer}}]{Jackson_06_GRS}
{Jackson}, J.~M., {Rathborne}, J.~M., {Shah}, R.~Y., {et~al.} 2006, \apjs, 163,
  145

\bibitem[{{Jiang} {et~al.}(2006){Jiang}, {Gao}, {Omont}, {Schuller}, \&
  {Simon}}]{Jiang_06_7-15ext}
{Jiang}, B.~W., {Gao}, J., {Omont}, A., {Schuller}, F., \& {Simon}, G. 2006,
  \aap, 446, 551

\bibitem[{{Jiang} {et~al.}(2003){Jiang}, {Omont}, {Ganesh}, {Simon}, \&
  {Schuller}}]{Jiang_03_extlaw}
{Jiang}, B.~W., {Omont}, A., {Ganesh}, S., {Simon}, G., \& {Schuller}, F. 2003,
  \aap, 400, 903

\bibitem[{{Li} \& {Greenberg}(2003)}]{Li_03_dustoverview}
{Li}, A. \& {Greenberg}, J.~M. 2003, in Solid State Astrochemistry, ed.
  V.~{Pirronello}, J.~{Krelowski}, \& G.~{Manic{\`o}}, 37--84

\bibitem[{{L{\'o}pez-Corredoira} {et~al.}(2002){L{\'o}pez-Corredoira},
  {Cabrera-Lavers}, {Garz{\'o}n}, \& {Hammersley}}]{L-C_02_MWscales}
{L{\'o}pez-Corredoira}, M., {Cabrera-Lavers}, A., {Garz{\'o}n}, F., \&
  {Hammersley}, P.~L. 2002, \aap, 394, 883

\bibitem[{{Lutz} {et~al.}(1996){Lutz}, {Feuchtgruber}, {Genzel}, {Kunze},
  {Rigopoulou}, {Spoon}, {Wright}, {Egami}, {Katterloher}, {Sturm},
  {Wieprecht}, {Sternberg}, {Moorwood}, \& {de Graauw}}]{Lutz_96_galcenter}
{Lutz}, D., {Feuchtgruber}, H., {Genzel}, R., {et~al.} 1996, \aap, 315, L269

\bibitem[{{Madsen} \& {Reynolds}(2005)}]{Madsen_05_diffuseISM}
{Madsen}, G.~J. \& {Reynolds}, R.~J. 2005, \apj, 630, 925

\bibitem[{{Mathis}(1990)}]{Mathis_90_dustreview}
{Mathis}, J.~S. 1990, \araa, 28, 37

\bibitem[{{McClure}(2009)}]{McClure_08_extlaw}
{McClure}, M. 2009, \apjl, 693, L81

\bibitem[{{Misiriotis} {et~al.}(2006){Misiriotis}, {Xilouris},
  {Papamastorakis}, {Boumis}, \& {Goudis}}]{Misiriotis_06_ISMmodel}
{Misiriotis}, A., {Xilouris}, E.~M., {Papamastorakis}, J., {Boumis}, P., \&
  {Goudis}, C.~D. 2006, \aap, 459, 113

\bibitem[{{Moore} {et~al.}(2005){Moore}, {Lumsden}, {Ridge}, \&
  {Puxley}}]{Moore_05_extlaw}
{Moore}, T.~J.~T., {Lumsden}, S.~L., {Ridge}, N.~A., \& {Puxley}, P.~J. 2005,
  \mnras, 359, 589

\bibitem[{{Nishiyama} {et~al.}(2006){Nishiyama}, {Nagata}, {Kusakabe},
  {Matsunaga}, {Naoi}, {Kato}, {Nagashima}, {Sugitani}, {Tamura}, {Tanab{\'e}},
  \& {Sato}}]{Nishi_06_extlaw}
{Nishiyama}, S., {Nagata}, T., {Kusakabe}, N., {et~al.} 2006, \apj, 638, 839

\bibitem[{{Nishiyama} {et~al.}(2009){Nishiyama}, {Tamura}, {Hatano}, {Kato},
  {Tanabe}, {Sugitani}, \& {Nagata}}]{Nishi_09_extlaw}
{Nishiyama}, S., {Tamura}, M., {Hatano}, H., {et~al.} 2009, ArXiv e-prints

\bibitem[{{Pedicelli} {et~al.}(2009){Pedicelli}, {Bono}, {Lemasle}, {Francois},
  {Groenewegen}, {Lub}, {Pel}, {Laney}, {Piersimoni}, {Romaniello}, {Buonanno},
  {Caputo}, {Cassisi}, {Castelli}, {Leurini}, {Pietrinferni}, {Primas}, \&
  {Pritchard}}]{Pedicelli_09_d-feh}
{Pedicelli}, S., {Bono}, G., {Lemasle}, B., {et~al.} 2009, ArXiv e-prints

\bibitem[{{Pineda} {et~al.}(2008){Pineda}, {Caselli}, \&
  {Goodman}}]{Pineda_08_COisotop}
{Pineda}, J.~E., {Caselli}, P., \& {Goodman}, A.~A. 2008, \apj, 679, 481

\bibitem[{{Rieke} \& {Lebofsky}(1985)}]{RiekeLeb_85_extlaw}
{Rieke}, G.~H. \& {Lebofsky}, M.~J. 1985, \apj, 288, 618

\bibitem[{{Rom{\'a}n-Z{\'u}{\~n}iga} {et~al.}(2007){Rom{\'a}n-Z{\'u}{\~n}iga},
  {Lada}, {Muench}, \& {Alves}}]{RZ_07_extlaw}
{Rom{\'a}n-Z{\'u}{\~n}iga}, C.~G., {Lada}, C.~J., {Muench}, A., \& {Alves},
  J.~F. 2007, \apj, 664, 357

\bibitem[{{Rosenthal} {et~al.}(2000){Rosenthal}, {Bertoldi}, \&
  {Drapatz}}]{Rosenthal_00_H2extlaw}
{Rosenthal}, D., {Bertoldi}, F., \& {Drapatz}, S. 2000, \aap, 356, 705

\bibitem[{{Skrutskie} {et~al.}(2006){Skrutskie}, {Cutri}, {Stiening},
  {Weinberg}, {Schneider}, {Carpenter}, {Beichman}, {Capps}, {Chester},
  {Elias}, {Huchra}, {Liebert}, {Lonsdale}, {Monet}, {Price}, {Seitzer},
  {Jarrett}, {Kirkpatrick}, {Gizis}, {Howard}, {Evans}, {Fowler}, {Fullmer},
  {Hurt}, {Light}, {Kopan}, {Marsh}, {McCallon}, {Tam}, {Van Dyk}, \&
  {Wheelock}}]{Skrutskie_06_2mass}
{Skrutskie}, M.~F., {Cutri}, R.~M., {Stiening}, R., {et~al.} 2006, \aj, 131,
  1163

\bibitem[{{Speck} {et~al.}(2009){Speck}, {Corman}, {Wakeman}, {Wheeler}, \&
  {Thompson}}]{Speck_09_dustformCstars}
{Speck}, A.~K., {Corman}, A.~B., {Wakeman}, K., {Wheeler}, C.~H., \&
  {Thompson}, G. 2009, \apj, 691, 1202

\bibitem[{{Weingartner} \& {Draine}(2001)}]{WeinDraine_01_dustsize}
{Weingartner}, J.~C. \& {Draine}, B.~T. 2001, \apj, 548, 296

\bibitem[{{Whittet}(2003)}]{Whittet_03_dust}
{Whittet}, D.~C.~B. 2003, {Dust in the Galactic Environment}, 2nd edn., Series
  in Astronomy and Astrophysics (Institute of Physics), pp. 66--109

\bibitem[{{Zhukovska} \& {Gail}(2009)}]{Zhu_09_dustevolve}
{Zhukovska}, S. \& {Gail}, H.-P. 2009, in The Evolving ISM in the Milky Way and
  Nearby Galaxies

\end{thebibliography}



\end{document}